\documentclass[paper,notoc]{JHEP3}
\usepackage{cite}
\usepackage{amsmath,amsfonts,amsbsy,amssymb} 
\usepackage{epsfig}
\usepackage{graphicx}

\newcommand\new{\newcommand}         

\newcommand{\s}{\ensuremath{\;}}

\newcommand{\mh}{\ensuremath{m_\mathrm{H}}}

\new{\emem}{{\ifmmode\mathrm{e}^-\else e$^-$\fi}}
\new{\epem}{{\ifmmode\mathrm{e}^+\else e$^+$\fi}}
\new{\zbo}  {{\ifmmode\mathrm{Z}\else Z\fi}}
\new{\wpm} {{\ifmmode\mathrm{W}^\pm\else W$^\pm$\fi}}
\new{\wbo} {{\ifmmode\mathrm{W}\else W\fi}}
\new{\epm} {{\ifmmode\mathrm{e^+e^-}\else $\mathrm{e^+e^-}$\fi}}
\new{\qq}  {{\ifmmode\mathrm{q}\else q\fi}}
\new{\qqb} {{\ifmmode\bar{\mathrm{q}}\else $\bar{\mathrm{q}}$\fi}}
\new{\tq}  {{\ifmmode\mathrm{t}\else t\fi}}
\new{\tqb} {{\ifmmode\bar{\mathrm{t}}\else $\bar{\mathrm{t}}$\fi}}
\new{\bq}  {{\ifmmode\mathrm{b}\else b\fi}}
\new{\bqb} {{\ifmmode\bar{\mathrm{b}}\else $\bar{\mathrm{b}}$\fi}}
\new{\ttbar}{\tq\tqb}
\new{\qqbar}{\qq\qqb}
\new{\gu}  {{\ifmmode\mathrm{g}\else g\fi}}
\new{\qqbarg}{\qq\qqb\gu}
\new{\pp}  {{\ifmmode\mathrm{p}\else p\fi}}

\new{\hh}  {{\ifmmode h\else $h$\fi}}
\new{\HH}  {{\ifmmode \mathrm{H}\else $\mathrm{H}$\fi}}
\new{\fe}  {{\ifmmode f\else $f$\fi}}
\new{\lp}  {{\ifmmode \ell\else $\ell$\fi}}
\new{\XX}  {{\ifmmode X\else $X$\fi}}
\new{\Vp}  {{\ifmmode V\else $V$\fi}}
\new{\Kzs} {{\ifmmode\mathrm{K}_\mathrm{S}^0\else $\mathrm{K}_\mathrm{S}^0$\fi}}
\new{\Kzl} {{\ifmmode\mathrm{K}_\mathrm{L}^0\else $\mathrm{K}_\mathrm{L}^0$\fi}}
\new{\Kp} {{\ifmmode\mathrm{K}\else $\mathrm{K}$\fi}}
\new{\ppHWW} {{\ifmmode\pp\pp\rightarrow\HH\rightarrow\wbo\wbo
                             \else $\pp\pp\rightarrow\HH\rightarrow\wbo\wbo$\fi}}
\new{\ppHWWlept} {{\ifmmode\pp\pp\rightarrow\HH\rightarrow\wbo\wbo\rightarrow\lp\nu\lp\nu
                             \else $\pp\pp\rightarrow\HH\rightarrow\wbo\wbo\rightarrow\lp\nu\lp\nu$\fi}}
\new{\ppHWWleptX} {{\ifmmode\pp\pp\rightarrow\HH+X\rightarrow\wbo\wbo+X\rightarrow\lp\nu\lp\nu+X
                             \else $\pp\pp\rightarrow\HH+X\rightarrow\wbo\wbo+X\rightarrow\lp\nu\lp\nu+X$\fi}}

\new{\HWWlept} {{\ifmmode\HH\rightarrow\wbo\wbo\rightarrow\lp\nu\lp\nu
                             \else $\HH\rightarrow\wbo\wbo\rightarrow\lp\nu\lp\nu$\fi}}
\new{\HWW} {{\ifmmode\HH\rightarrow\wbo\wbo
                             \else $\HH\rightarrow\wbo\wbo$\fi}}

\new{\WW} {{\ifmmode\wbo\wbo
                             \else $\wbo\wbo$\fi}}
                            
\new{\pptt}{{\ifmmode\pp\pp\rightarrow\ttbar
                             \else $\pp\pp\rightarrow\ttbar$\fi}}
\new{\ppWW}{{\ifmmode\pp\pp\rightarrow\wbo\wbo
                             \else $\pp\pp\rightarrow\wbo\wbo$\fi}}
\new{\Hgammagamma} {{\ifmmode\HH\rightarrow\gamma\gamma
                             \else $\HH\rightarrow\gamma\gamma$\fi}}

\new{\LEP}        {\mbox{\small\textsc{LEP}}}
\new{\LEPONE}     {\mbox{\small\textsc{LEP1}}}
\new{\LEPTWO}     {\mbox{\small\textsc{LEP2}}}
\new{\CERN}       {\mbox{\small\textsc{CERN}}}
\new{\ALEPH}      {\mbox{\small\textsc{ALEPH}}}
\new{\DELPHI}     {\mbox{\small\textsc{DELPHI}}}
\new{\LD}         {\mbox{\small\textsc{L3}}}
\new{\OPAL}       {\mbox{\small\textsc{OPAL}}}
\new{\SPS}        {\mbox{\small\textsc{SPS}}}
\new{\TEVATRON}   {\mbox{\small\textsc{TEVATRON}}}
\new{\LHC}        {\mbox{\small\textsc{LHC}}}
\new{\FERMILAB}   {\mbox{\small\textsc{FERMILAB}}}
\new{\CDF}        {\mbox{\small\textsc{CDF}}}
\new{\DZERO}      {\mbox{\small\textsc{D0}}}
\new{\CTEQ}        {\mbox{\small\textsc{CTEQ}}}
\new{\FNAL}        {\mbox{\small\textsc{FNAL}}}
\new{\ATLAS}        {\mbox{\small\textsc{ATLAS}}}
\new{\CMS}        {\mbox{\small\textsc{CMS}}}

\new{\eV}         {{\ifmmode {\mathrm{ eV}}\else ${\mathrm{ eV}}$\fi}}
\new{\MeV}        {{\ifmmode {\mathrm{ MeV}}\else ${\mathrm{ MeV}}$\fi}}
\new{\MeVc}       {{\ifmmode {\mathrm{ MeV}}/c\else ${\mathrm{ MeV}}/c$\fi}}
\new{\MeVcc}      {{\ifmmode {\mathrm{ MeV}}/c^2\else ${\mathrm{ MeV}}/c^2$\fi}}
\new{\GeV}        {{\ifmmode {\mathrm{ GeV}}\else ${\mathrm{ GeV}}$\fi}}
\new{\GeVc}       {{\ifmmode {\mathrm{ GeV}}/c\else ${\mathrm{GeV}}/c$\fi}}
\new{\GeVcc}      {{\ifmmode {\mathrm{ GeV}}/c^2\else ${\mathrm{GeV}}/c^2$\fi}}
\new{\TeV}        {{\ifmmode {\mathrm{ TeV}}\else ${\mathrm{ TeV}}$\fi}}

\new{\fb}        {{\ifmmode {\mathrm{ fb}}\else ${\mathrm{ fb}}$\fi}}
\new{\fbinv}   {{\ifmmode {\mathrm{ fb}^{-1}}\else ${\mathrm{ fb}^{-1}}$\fi}}
\new{\pb}        {{\ifmmode {\mathrm{ pb}}\else ${\mathrm{ pb}}$\fi}}
\new{\pbinv}   {{\ifmmode {\mathrm{ pb}^{-1}}\else ${\mathrm{ pb}^{-1}}$\fi}}

\new{\JS}         {\mbox{\small\textsc{JETSET}}}
\new{\HERWIG}         {\mbox{\small\textsc{HERWIG}}}
\new{\PYTHIA}         {\mbox{\small\textsc{PYTHIA}}}
\new{\AR}         {\mbox{\small\textsc{ARIADNE}}}
\new{\PY}         {\mbox{\small\textsc{PYTHIA}}}
\new{\JSv}        {\mbox{\small\textsc{JETSET\ 7.405}}}
\new{\HWo}        {\mbox{\small\textsc{HERWIG\ 5.8}}}
\new{\HWn}        {\mbox{\small\textsc{HERWIG\ 5.9}}}
\new{\ARv}        {\mbox{\small\textsc{ARIADNE\ 4.05}}}
\new{\PYv}        {\mbox{\small\textsc{PYTHIA\ 5.7}}}

\new{\fehip}        {\mbox{\small\textsc{FEHiP}}}
\new{\fewz}        {\mbox{\small\textsc{FEWZ}}}
\new{\hqt}        {\mbox{\small\textsc{HqT}}}
\new{\mcnlo}        {\mbox{\small\textsc{MC@NLO}}}
\new{\pvegas}        {\mbox{\small\textsc{PVEGAS}}}

\new{\Mz}         {{\ifmmode M_{\mathrm{ Z}}
                    \else $M_{\mathrm{ Z}}$\fi}}
\new{\Mzsq}       {{\ifmmode M^2_{\mathrm{ Z}}
                    \else $M^2_{\mathrm{ Z}}$\fi}}
\new{\Mw}         {{\ifmmode M_{\mathrm{ W}}
                    \else $M_{\mathrm{ W}}$\fi}}
                    
\new{\MH}         {{\ifmmode m_{\mathrm{ H}}
                    \else $m_{\mathrm{ H}}$\fi}}

\new{\as}[1]      {{\ifmmode\alpha^{#1}_s
                    \else$\alpha^{#1}_s$\fi}}
\new{\asx}[1]      {{\ifmmode a^{#1}_s
                    \else $a^{#1}_s$\fi}}
\new{\asb}[1]     {{\ifmmode\overline{\alpha}^{#1}_s
                    \else $\overline{\alpha}^{#1}_s$\fi}}
\new{\asmz}       {{\ifmmode\alpha_s(\Mzsq)
                    \else $\alpha_s(\Mzsq)$\fi}}
\new{\lqcd}       {{\ifmmode\Lambda_{\mathrm{ QCD}}
                    \else $\Lambda_{\mathrm{ QCD}}$\fi}}
\new{\lqcdsq}     {{\ifmmode\Lambda^2_{\mathrm{ QCD}}
                    \else $\Lambda^2_{\mathrm{ QCD}}$\fi}}
\new{\llla}       {{\ifmmode\Lambda_{\mathrm{ LLA}}
                    \else $\Lambda_{\mathrm{ LLA}}$\fi}} 
\new{\lmsbar}[1]  {{\ifmmode \Lambda^{(#1)}_{\overline{\mathrm{MS}}}
                    \else $\Lambda^{(#1)}_{\overline{\mathrm{MS}}}$\fi}}
\new{\lmsb}       {{\ifmmode \Lambda_{\overline{\mathrm{MS}}}
                    \else $\Lambda_{\overline{\mathrm{MS}}}$\fi}}
\new{\lmsbsq}     {{\ifmmode \Lambda^{2}_{\overline{\mathrm{MS}}}
                    \else $\Lambda^{2}_{\overline{\mathrm{MS}}}$\fi}}
\new{\pt}       {{\ifmmode p_{\mathrm{T}}
                    \else $p_{\mathrm{T}}$\fi}}
\new{\etmisscut}       {{\ifmmode E_{\mathrm{T}}^{\mathrm{miss,cut}}
                    \else $E_{\mathrm{T}}^{\mathrm{miss,cut}}$\fi}}
\new{\ptlmin}       {{\ifmmode p_{\mathrm{T}}^{\lp\mathrm{min}}
                    \else $p_{\mathrm{T}}^{\lp\mathrm{min}}$\fi}}
\new{\ptlmaxcut}       {{\ifmmode \mathrm{p}_{\mathrm{T,max}}
                    \else $\mathrm{p}_{\mathrm{t,max}}^{\mathrm{cut}}$\fi}} 
\new{\ptlep}       {{\ifmmode p_{\mathrm{T}}^{\mathrm{lepton}}
                    \else $p_{\mathrm{T}}^{\mathrm{lepton}}$\fi}}  
\new{\ptveto}       {{\ifmmode p_{\mathrm{T}}^{\mathrm{veto}}
                    \else $p_{\mathrm{T}}^{\mathrm{veto}}$\fi}}          
\new{\kt}       {{\ifmmode k_{\mathrm{T}}
                    \else $k_{\mathrm{T}}$\fi}}          

\newcommand{\eps}{\epsilon}
\newcommand{\mhiggs}{m_H}
\def\Bubble(#1,#2){\mathrm{B}\left(#1,#2\right)}
\def\TrianB(#1,#2){\mathrm{C}\left(#1,#2\right)}
\def\TrianA(#1,#2,#3){\mathrm{C}_1\left(#1,#2,#3\right)}
\def\Boxx(#1,#2,#3){\mathrm{D}\left(#1,#2,#3\right)}
\def\s(#1,#2){s_{#1 #2}}
\def\half{\frac{1}{2}}

\title{\boldmath HPro: A NLO Monte-Carlo for Higgs production via gluon fusion 
with finite heavy quark masses}

\author{Charalampos Anastasiou\\
  Institute for Theoretical Physics, ETH Zurich,\\
  8093 Zurich, Switzerland\\
  E-mail: \email{babis@phys.ethz.ch}}
\author{Stefan Bucherer\\
  Institute for Theoretical Physics, ETH Zurich,\\
  8093 Zurich, Switzerland\\
  E-mail: \email{stefabu@itp.phys.ethz.ch}}
\author{Zoltan Kunszt\\
  Institute for Theoretical Physics, ETH Zurich,\\
  8093 Zurich, Switzerland\\
  E-mail: \email{kunszt@itp.phys.ethz.ch}}

\abstract{ 
We compute fully differential next-to-leading order QCD cross-sections  for 
Higgs boson production via gluon fusion in the Standard Model.  We maintain 
the full dependence of the cross-sections on the top and bottom quark mass.  
We find that finite quark mass effects are important given the achieved 
precision of QCD  predictions for  gluon fusion. 
Our Monte-Carlo program {\tt HPro} can correct existing NNLO fully differential calculations, which employ the 
approximation of an infinitely  heavy top and a vanishing bottom quark Yukawa
coupling, for heavy quark finite mass 
effects through NLO. 
} 

\begin{document}
\section{Introduction}
\label{sec:intro}

QCD  radiative effects in the gluon fusion process for the production of Higgs bosons 
at hadron colliders have been the subject of detailed theoretical investigations in the last 
two decades.  The next-to-leading order QCD corrections to the inclusive cross-section 
were computed already in the nineties~\cite{Dawson:1990zj,Djouadi:1991tka,Spira:1995rr}. 
In these works it was shown that QCD perturbative corrections are substantial. 

Exact calculations of perturbative corrections in the gluon fusion process, such as in Ref.~\cite{Spira:1995rr},  
are technically involved due to the presence of massive quark  loops already at the leading order. 
With these pioneering NLO computations, the quality of the simplifying
approximation of an infinitely heavy top quark and vanishing Yukawa couplings
for all other quarks  could be assessed. 

Such an approximation is indispensable for computing QCD perturbative corrections beyond NLO.
Next-to-next-to-leading order QCD corrections for the $gg \to H $ inclusive cross-section
were computed in Refs~\cite{Harlander:2002wh,Anastasiou:2002yz,Ravindran:2003um}. NNLO corrections, within the 
same approximation,  for fully differential cross-sections  were  computed in 
Refs~\cite{Anastasiou:2004xq,fehip,Anastasiou:2007mz} and in Refs~\cite{Catani:2007vq,hnnlo}. 

The  theoretical uncertainty due to scale  variations of  NNLO inclusive and differential cross-sections  is by now  
remarkably small. While existing tools, such as {\tt HIGLU}~\cite{Spira:1996if} can be used to correct the predictions for the total 
cross-section for finite quark-mass effects through NLO, fully differential cross-sections for the Higgs  boson and 
its decay products  cannot  be corrected for the same effects with existing  tools.  We remedy this situation in our 
publication. 

We have written a parton level Monte-Carlo program {\tt HPro} 
which computes fully differential cross-sections at NLO in QCD while keeping the exact dependence on  
the finite top and bottom quark mass. {\tt HPro} includes 
the decays of the  Higgs boson to photons and  four-lepton final states. It can 
be used  in conjunction with the fully differential 
NNLO program {\tt FEHiP}~\cite{fehip,Anastasiou:2007mz}, correcting for finite quark mass 
effects through NLO. An earlier version of {\tt HPro} has been used in
  \cite{Anastasiou:2008tj} to estimate the finite bottom mass effects. Later in a
  similar study~\cite{deFlorian:2009hc} these effects have been also accounted for using {\tt HIGLU}.

We  present  the calculation method  in Section~\ref{sec:method}. 
In Section~\ref{sec:total} we use {\tt HPro} to compute the NLO total cross-section and review the quality of the approximations 
which are usually made for heavy quark loops in  the gluon fusion process.  In Section~\ref{sec:differential}, we compute for illustration
various kinematic distributions and study the effect of  finite  quark masses to their  shape. 
We  present our conclusions in Section~\ref{sec:conclusion}. 

\section{Implementation}
\label{sec:method}

The computation of the NLO corrections to the gluon fusion process requires  the two-loop 
$gg \to h$ amplitude. This has been first computed by means of an one-dimensional 
integral representation derived in Ref.~\cite{Spira:1995rr}. 
In Ref.~\cite{Harlander:2005rq}, this  result was  expressed analytically in terms of harmonic 
polylogarithms using a ``series expansion and  matching'' method. Independent analytic evaluations 
of the two-loop amplitude were performed in Ref.~\cite{Anastasiou:2006hc}  and
in Ref.~\cite{Aglietti:2006tp}.

At NLO also real radiation sub-processes $gg\to gh$, $qg\to qh$ and
$q\bar{q}\to gh$ contribute. The corresponding matrix elements have been first
computed in Ref.~\cite{Ellis:1987xu}. In Ref.~\cite{Baur:1989cm} these matrix elements
have been expressed in terms of standard one-loop scalar integrals and
helicity amplitudes. We have recomputed these contributions along the lines of Ref.~\cite{Baur:1989cm} and found full
agreement. In particular we have also compared to a very
recent calculation in Ref.~\cite{Keung:2009bs}. For completeness we present the results in Appendix~\ref{sec:app_real}.

Virtual and real corrections develop singularities which only cancel in their
combination and by adding the appropriate collinear counter-term for PDF
evolution. In order to obtain an expression suited for numerical integration
we apply the FKS subtraction method \cite{Frixione:1995ms}. Recently this method has been applied in Ref.~\cite{Alioli:2008tz} for the matching
of NLO Higgs production to shower Monte Carlo in the heavy top mass approximation. 

We note that in the case of total cross sections the straightforward application of the FKS subtraction method
leads to the same analytic formula for the subtraction terms that has been
found in  Ref.~\cite{Spira:1995rr}. With appropriate insertion of 
measurement functions we easily get analytic expressions suited for the
calculation of any differential distribution.  In practice one might think of the
measurement function as a vector valued function with each component being a
bin of a certain distribution. There exist implementations of the VEGAS
algorithm supporting vector functions, such as the VEGAS routine in the CUBA
library \cite{Hahn:2004fe}, which allows to obtain a reliable Monte-Carlo
error estimate for each bin. In combination with dedicated decay routines for
the diphoton and leptonic final states, this feature enables us to produce
various distributions relevant for experimental searches of the Higgs boson,
corrected by finite mass effects, as we will demonstrate in the following
sections.

\section{Total cross-section at NLO}
\label{sec:total}

In this section, we revisit the gluon fusion cross-section  at NLO. This serves as a  check of our Monte-Carlo {\tt HPro} against the predictions of {\tt HIGLU}~\cite{Spira:1996if}, and to 
emphasize the importance of finite  quark mass effects in Higgs boson production.   
For the numerical results  of this paper we use MSTW 2008 parton distribution
functions~\cite{Martin:2009iq}.  

\begin{figure}[th]
  \begin{center}
\includegraphics[width=0.48\textwidth]{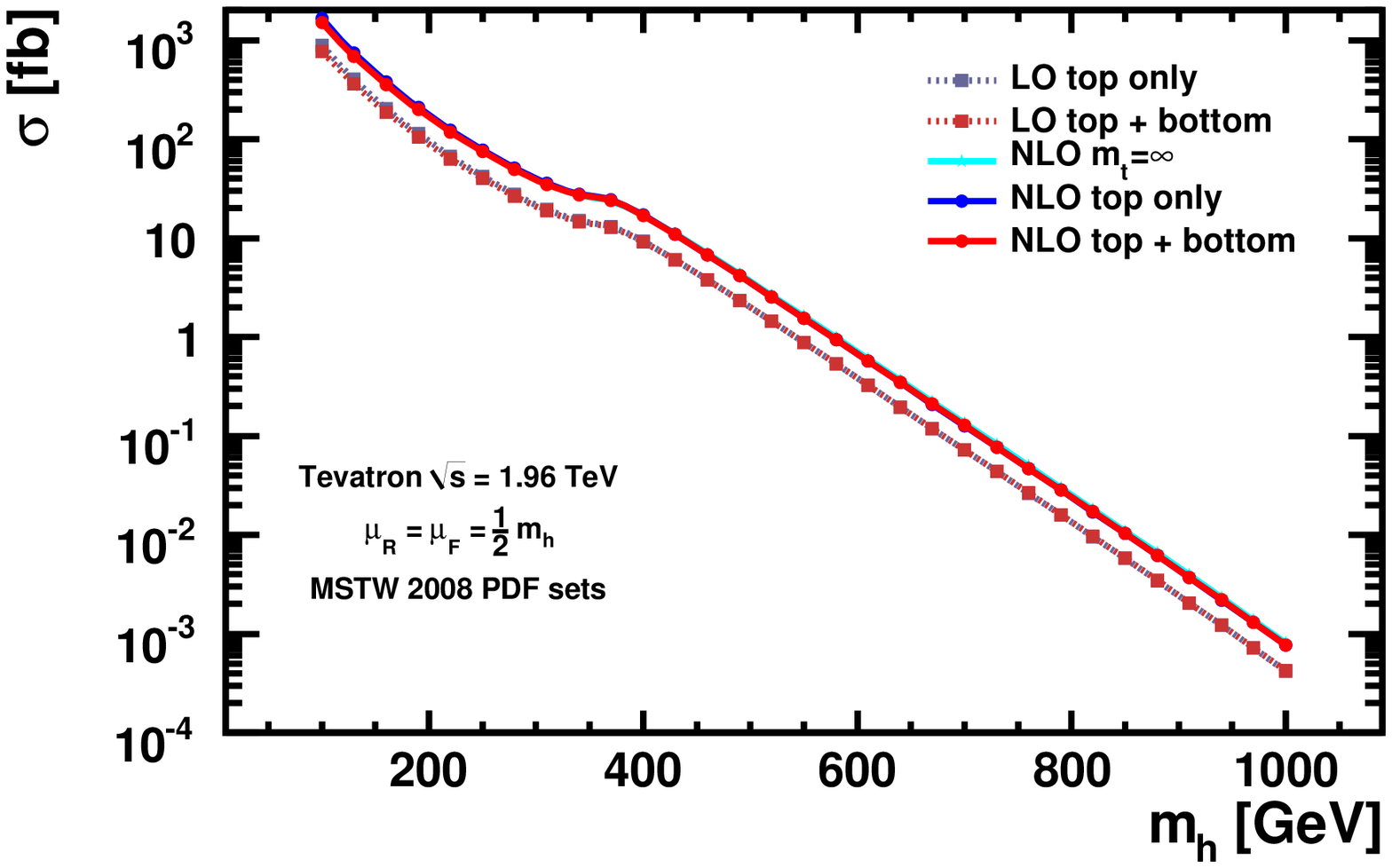}
\includegraphics[width=0.48\textwidth]{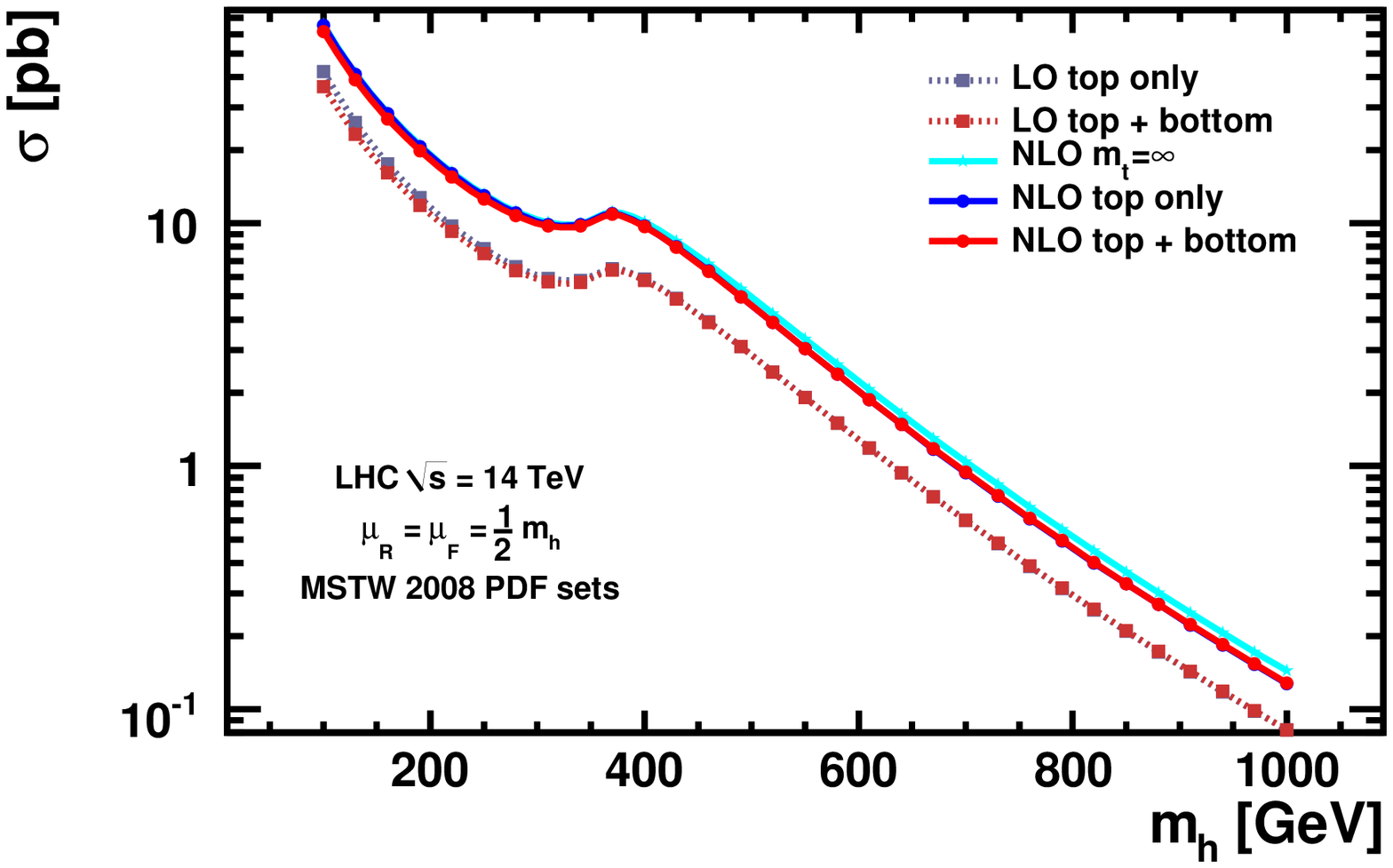}
  \end{center}
  \caption{\label{fig:sigma}
   Total Cross-Section at Tevatron and LHC.}
\end{figure}

We  begin our study by revisiting the total cross-section in the LO and NLO  approximation as a function of the  Higgs boson mass (Fig.~\ref{fig:sigma}). 
As it is well known~\cite{Dawson:1990zj,Djouadi:1991tka,Spira:1995rr},  NLO QCD perturbative corrections are  substantial. We  note  
here that the perturbative corrections  are slightly  smaller with the latest parton densities~\cite{Martin:2009iq},  mainly due  to the  higher value of $\alpha_s$ used 
at leading order. We also note that a  significant QCD correction is found at 
NNLO~\cite{Harlander:2002wh,Anastasiou:2002yz,Ravindran:2003um}, which is not 
included in the results of this  article.  NNLO corrections stabilize  the perturbative expansion and reduce the  scale variation to the $\sim 10\%$ level. However, NNLO computations rely on an approximate treatment of  heavy quark  loops. 

In Fig.~\ref{fig:sigma}, we show  the effects  of different treatments for the heavy 
quarks.  With {\tt HPro}, we compute the exact LO and NLO cross-sections, 
where all loop diagrams with massive top and bottom quarks  are evaluated  exactly (we denote with ``top+bottom'' the  corresponding cross-sections 
in the plots of this paper). An approximation which can be  made, 
is to consider a vanishing  bottom Yukawa coupling and to evaluate exactly only 
the top-quark loops; in our plots, we denote this  approximation as ``top-only''. 
NNLO computations  are performed in what is known as the infinitely heavy top-quark 
approximation. In the ``$m_{\rm top} = \infty$'' approximation, the bottom Yukawa 
coupling is set to zero, and the cross-section at higher orders is estimated by the 
formula,
\begin{equation}
\label{eq:hqet}
\sigma_{\rm (N)NLO}^{m_{\rm top} = \infty} = \sigma_{\rm LO}^{{\rm top-only}} \times 
\lim_{m_{\rm top} \to \infty} \left( 
\frac
{\sigma_{\rm (N)NLO}^{{\rm top-only}} } 
{\sigma_{\rm LO}^{{\rm top-only}} }\right), 
\end{equation}
where bottom quark loops are ignored, and the leading order  cross-section is  reweighted 
with the ratio of the cross-sections  at  higher orders and the leading order in the limit of 
an infinite  top-quark mass. In order to compare the different approximations
we introduce 
\begin{equation}
\delta X^{i} = \frac{X^{i} - X^{m_{\rm top}=\infty}}{X^{m_{\rm top}=\infty}}
\end{equation}
where $X$ is the inclusive cross section or a normalized differential cross
section and $i$ is labelling the contribution (top+bottom, top-only, etc.) 

\begin{figure}[th]
  \begin{center}
\includegraphics[width=0.48\textwidth]{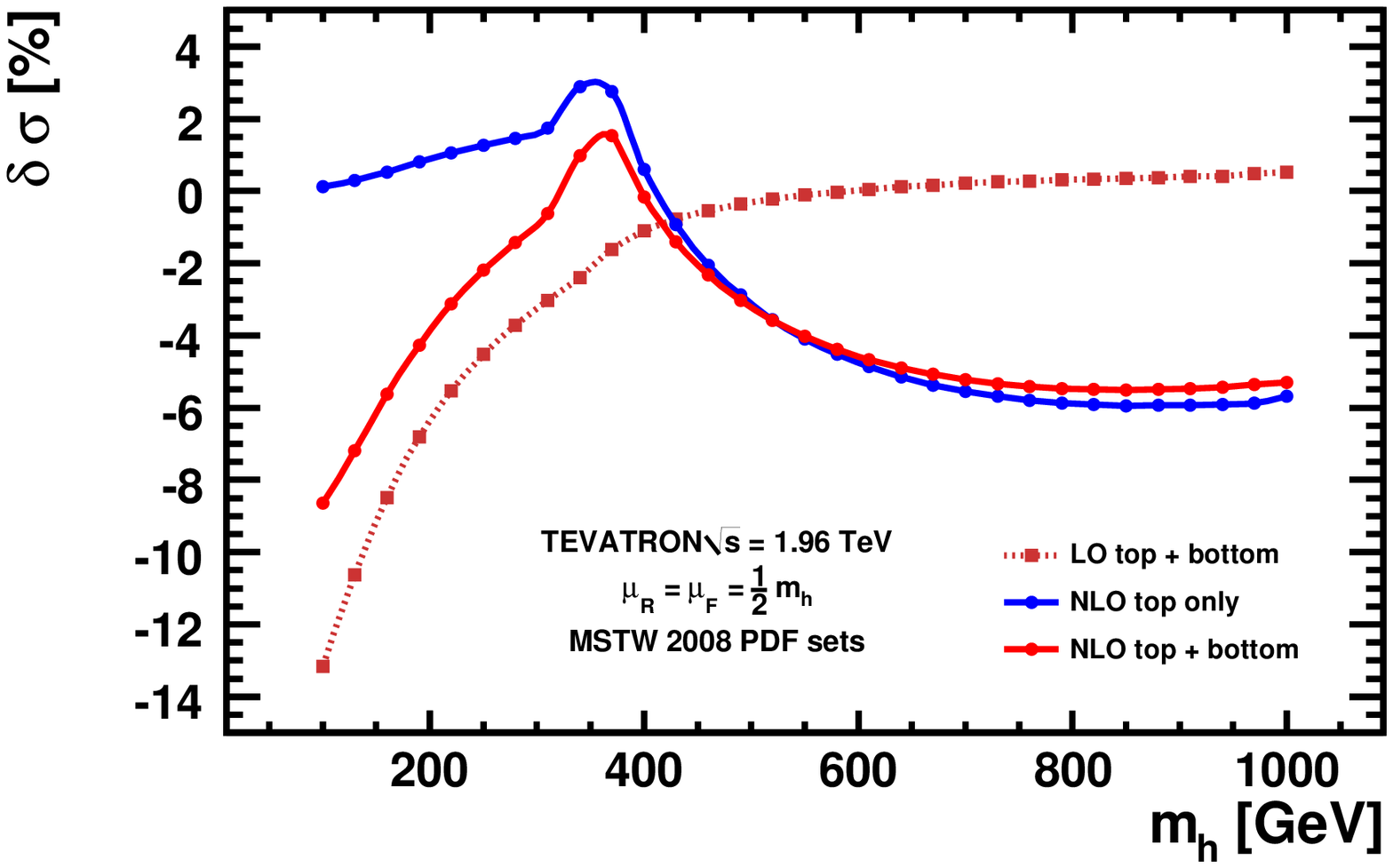}
\includegraphics[width=0.48\textwidth]{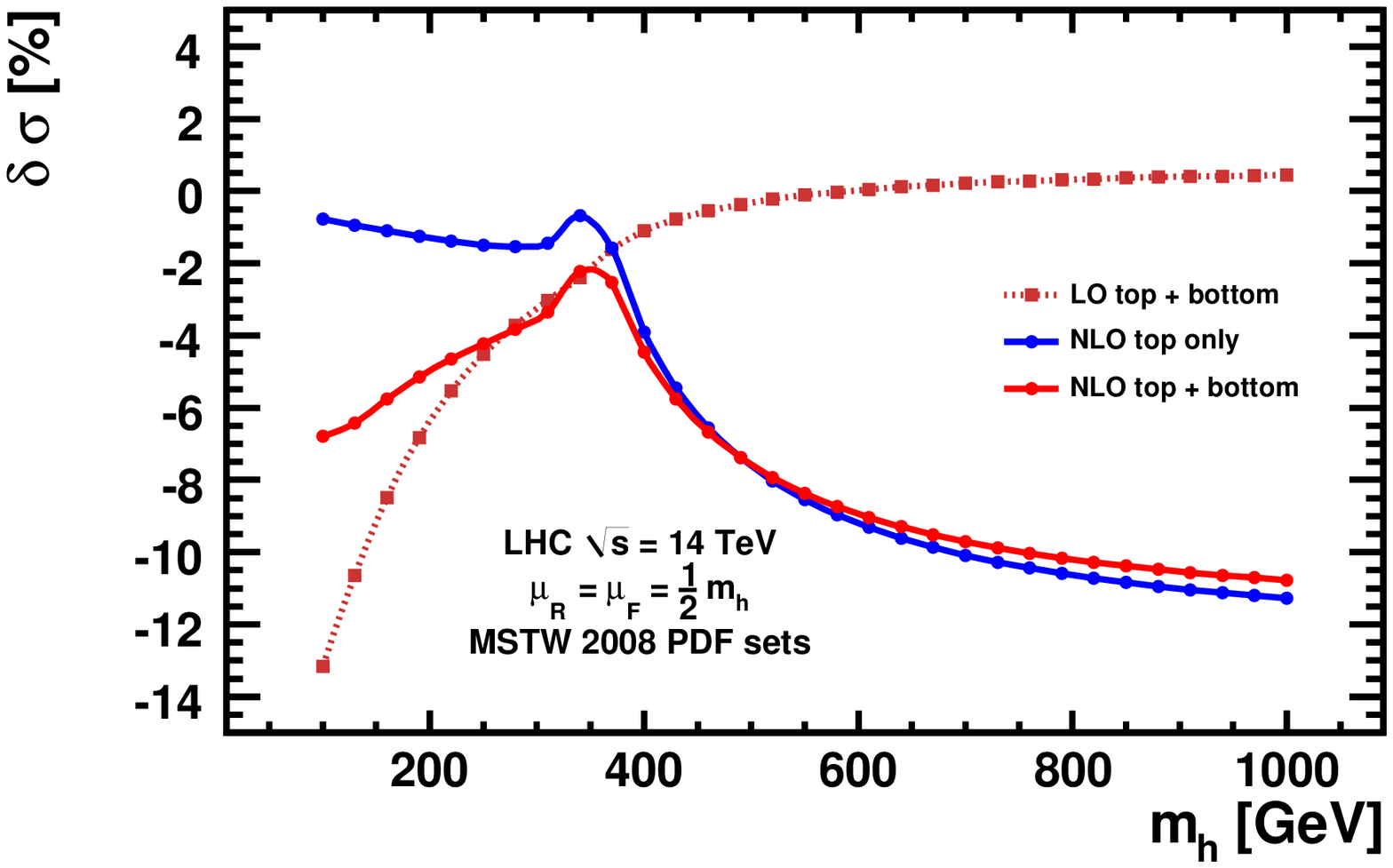}
  \end{center}
  \caption{\label{fig:sigma_lhc_diffs}
 Percent differences 
 of the exact NLO (LO) total cross-section with finite top bottom masses or
 the NLO (LO) total cross-section 
with exact top mass effects but  zero bottom Yukawa coupling with respect to the usual approximation at Tevatron and the LHC.}
\end{figure}  

In Fig.~\ref{fig:sigma_lhc_diffs}, we show the cross-section 
deviations  from the $m_{\rm top} = \infty$ approximation of Eq.~\ref{eq:hqet} when ``top-only'' (blue) and the  complete ``top-bottom'' mass effects are taken into account. 
``Top-only'' contributions  are approximated within a couple of  a  percent up
to the
$\mh = 2 m_{\rm top}$ threshold.  However, for  a light Higgs boson, bottom 
quark contributions are important and  can reach $\sim -8\%$.   It is then important that   bottom loops are taken into account for a  precise evaluation of  the total cross-section~\cite{Anastasiou:2008tj}. We also observe that  the contribution from  bottom-quark loops decreases at NLO in comparison to  LO.  

\begin{figure}[th]
  \begin{center}
  \includegraphics[width=0.48\textwidth]{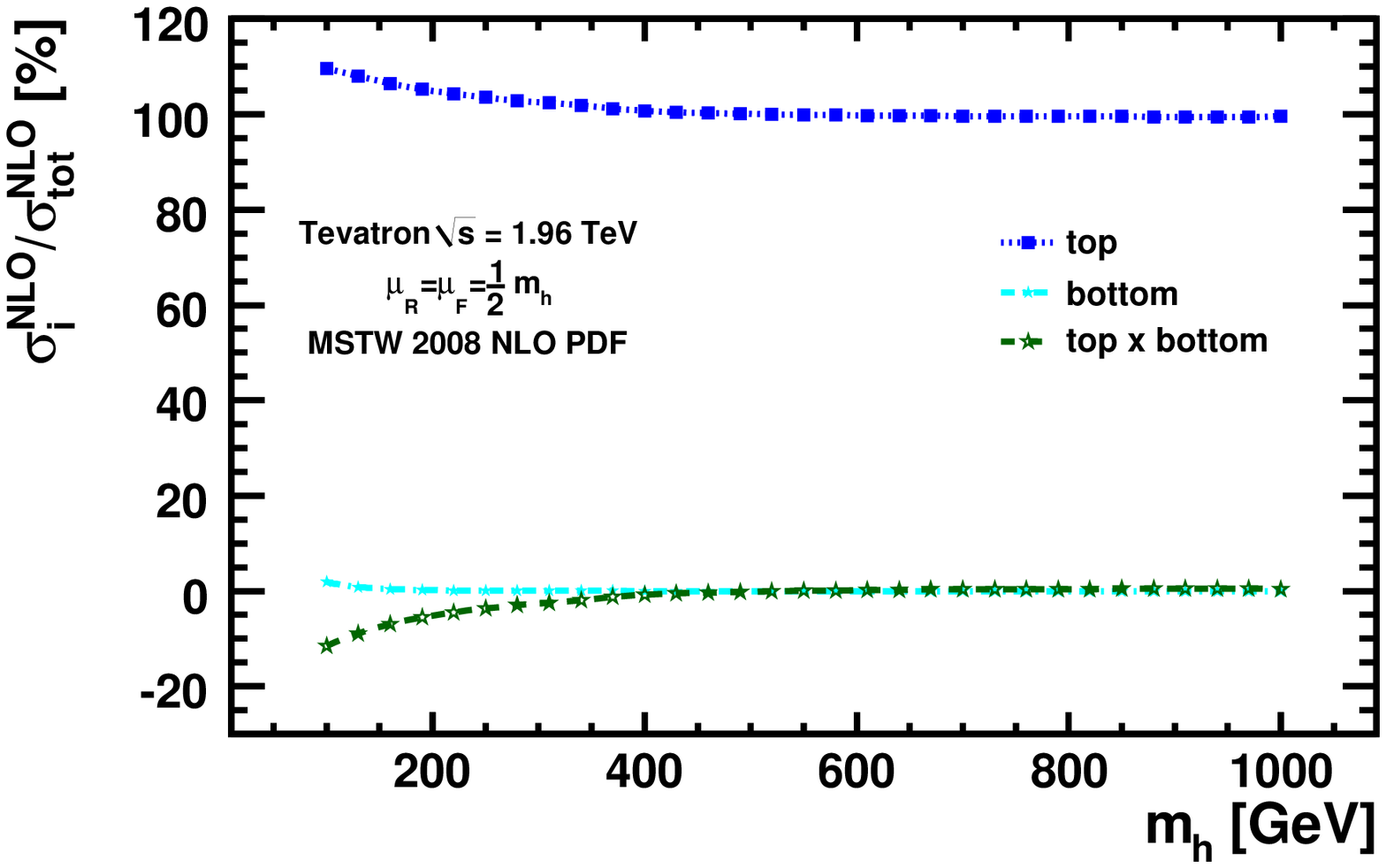}
  \includegraphics[width=0.48\textwidth]{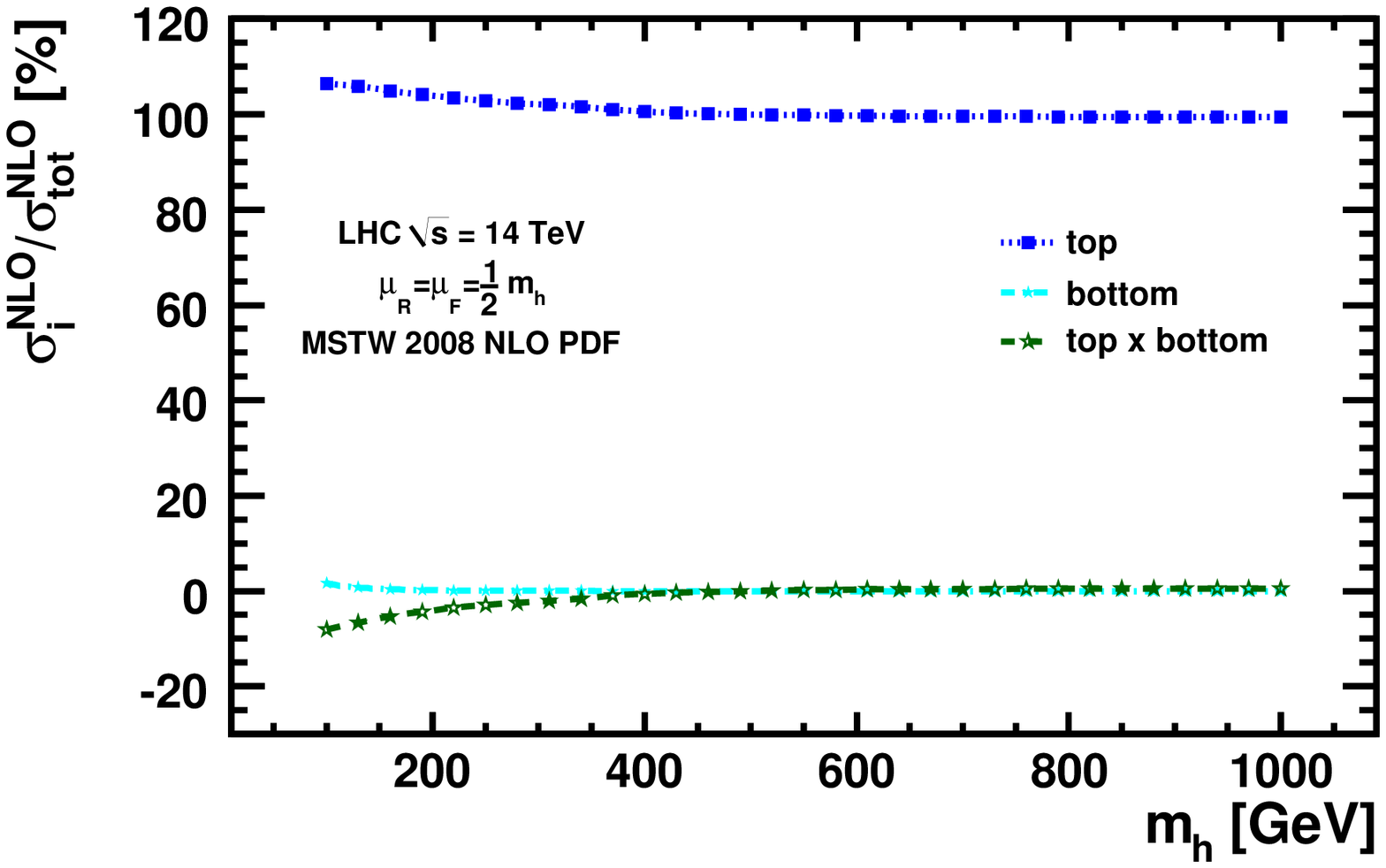}
  \end{center}
  \caption{\label{fig:top_vs_bot}
Ratio in percentage of top-only, bottom-only and top-bottom interference components with respect to the total cross-section at
NLO, for Tevatron and LHC.}
\end{figure}  

In Fig.~\ref{fig:top_vs_bot}, we show  the relative contributions to the NLO total cross-section 
from top-quark loops only,  bottom quark-loops only, and  from the interference of top and bottom loops.  
Top-only contributions  are dominant, while bottom-only contributions  are
negligible over the whole Higgs mass range.  
Top-bottom interference terms are important at the few percent level and are negative for a light Higgs boson.  
It should be  noted that  the relative importance  of the three contributions for a  heavy Higgs boson or a
pseudo-scalar Higgs boson in the  MSSM may be  drastically different than in the Standard Model~\cite{Langenegger:2006wu}. 

\begin{figure}[th]
  \begin{center}
 \includegraphics[width=0.48\textwidth]{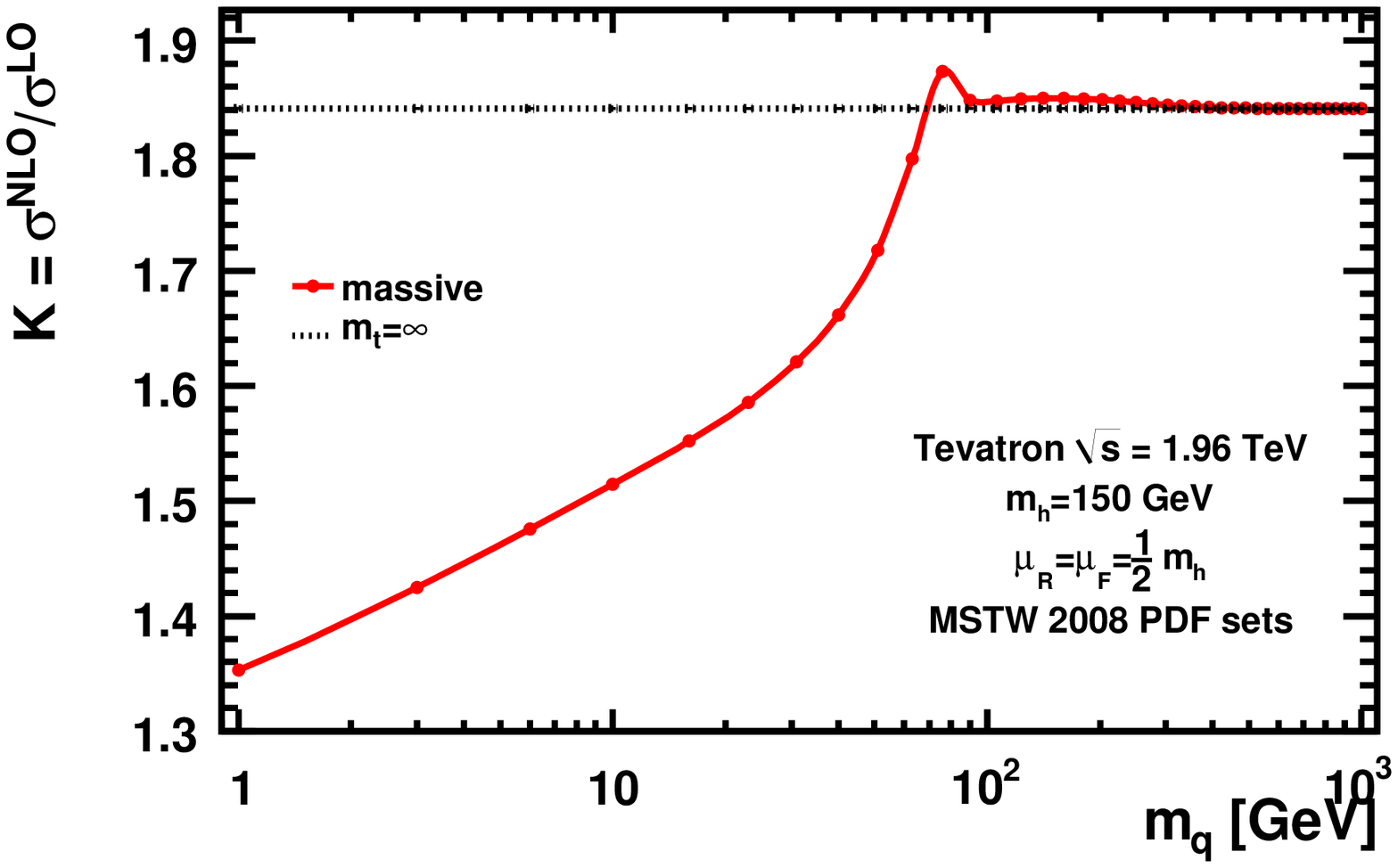}
 \includegraphics[width=0.48\textwidth]{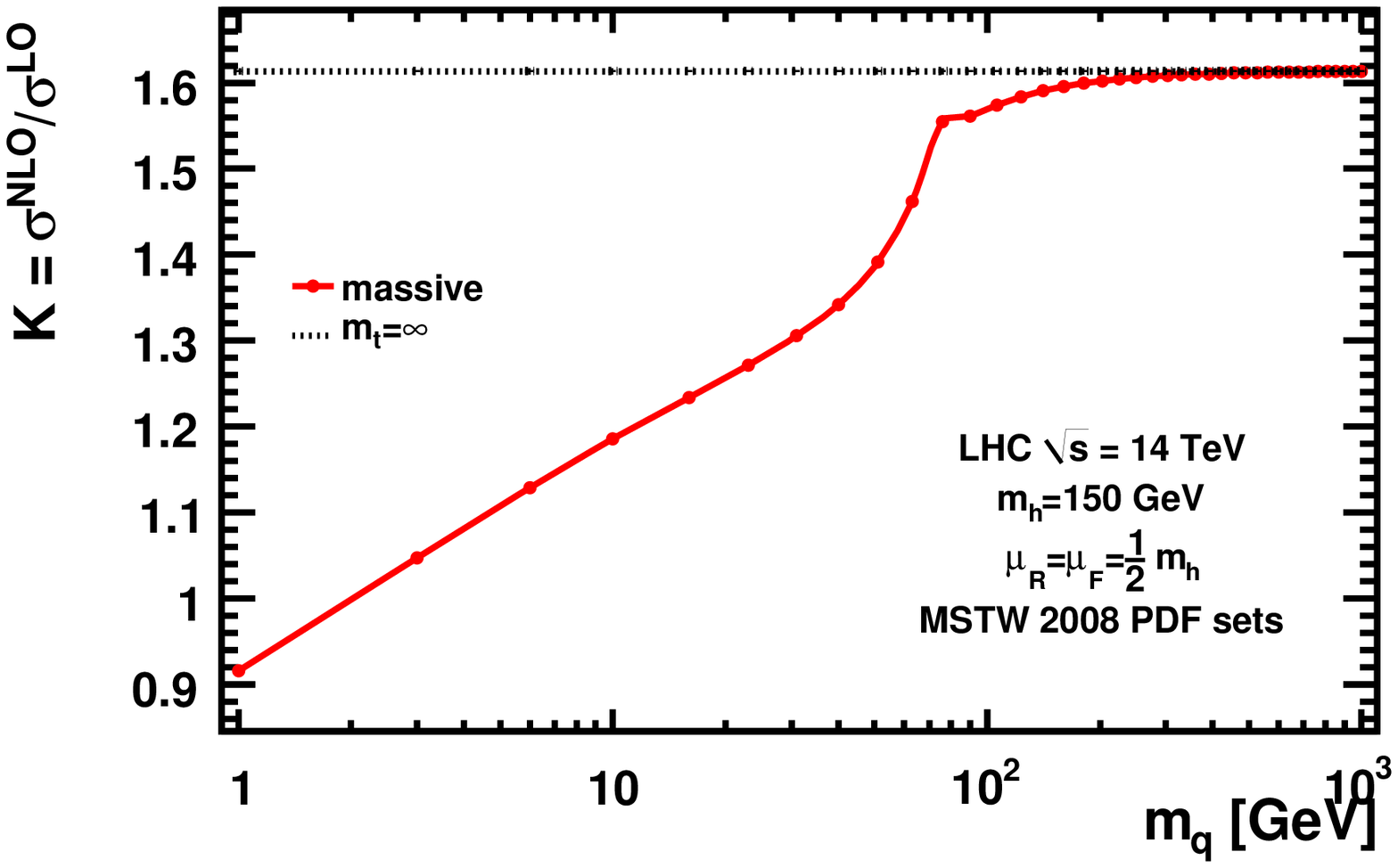}
 \includegraphics[width=0.48\textwidth]{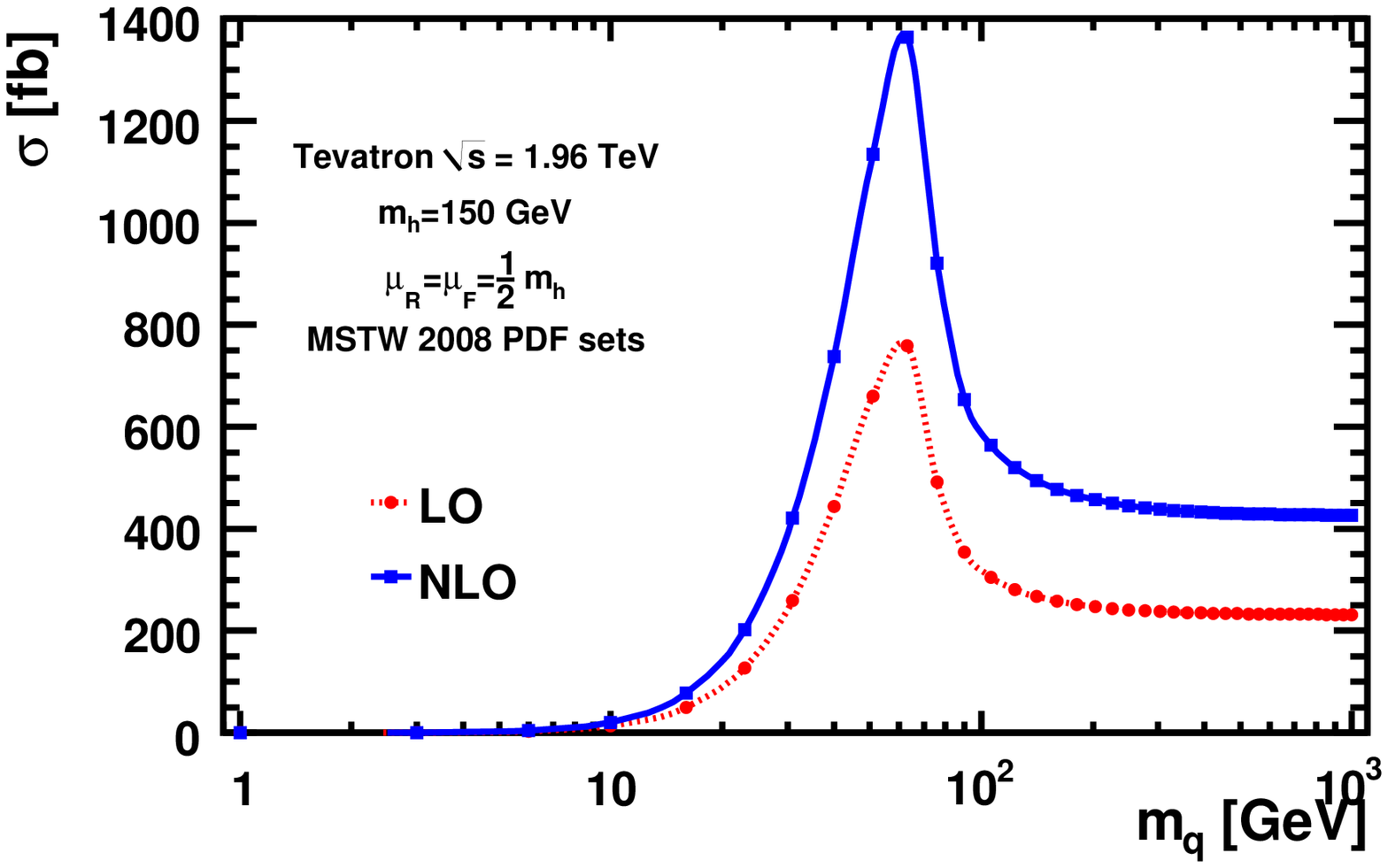} 
 \includegraphics[width=0.48\textwidth]{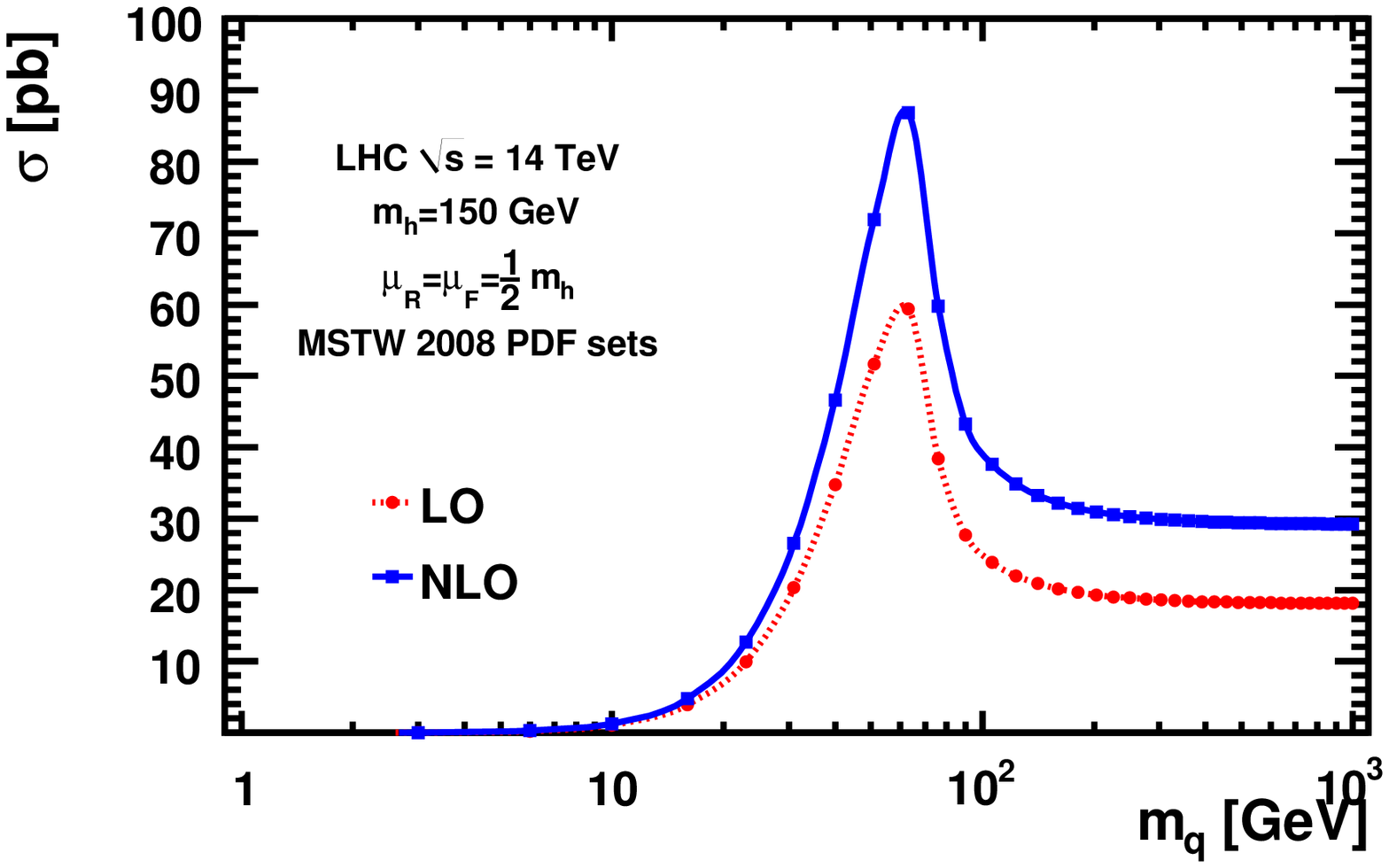}
\end{center}
  \caption{\label{fig:sigma_quark}
NLO cross-section and  K-factor  for  the  gluon fusion process via one only heavy quark at the  Tevatron and  the LHC, as a function of the quark mass.}
\end{figure}  
The magnitude of QCD corrections  depends  strongly on the 
mass value of the heavy quark in the loops if we use the pole mass for the
$q\bar{q}H$ coupling. 
In Fig.~\ref{fig:sigma_quark} we consider the cross-section for the gluon fusion cross-section at  the Tevatron and the 
LHC, considering only one  heavy quark with a mass $m_q$.  We find that,  in the pole scheme,  
 the K-factor is reduced  significantly for  small  values of the quark mass
 consistent with neglecting the running of the quark mass in the Higgs
 coupling at leading order. 
 
\begin{figure}[th]
  \begin{center}
 \includegraphics[width=0.48\textwidth]{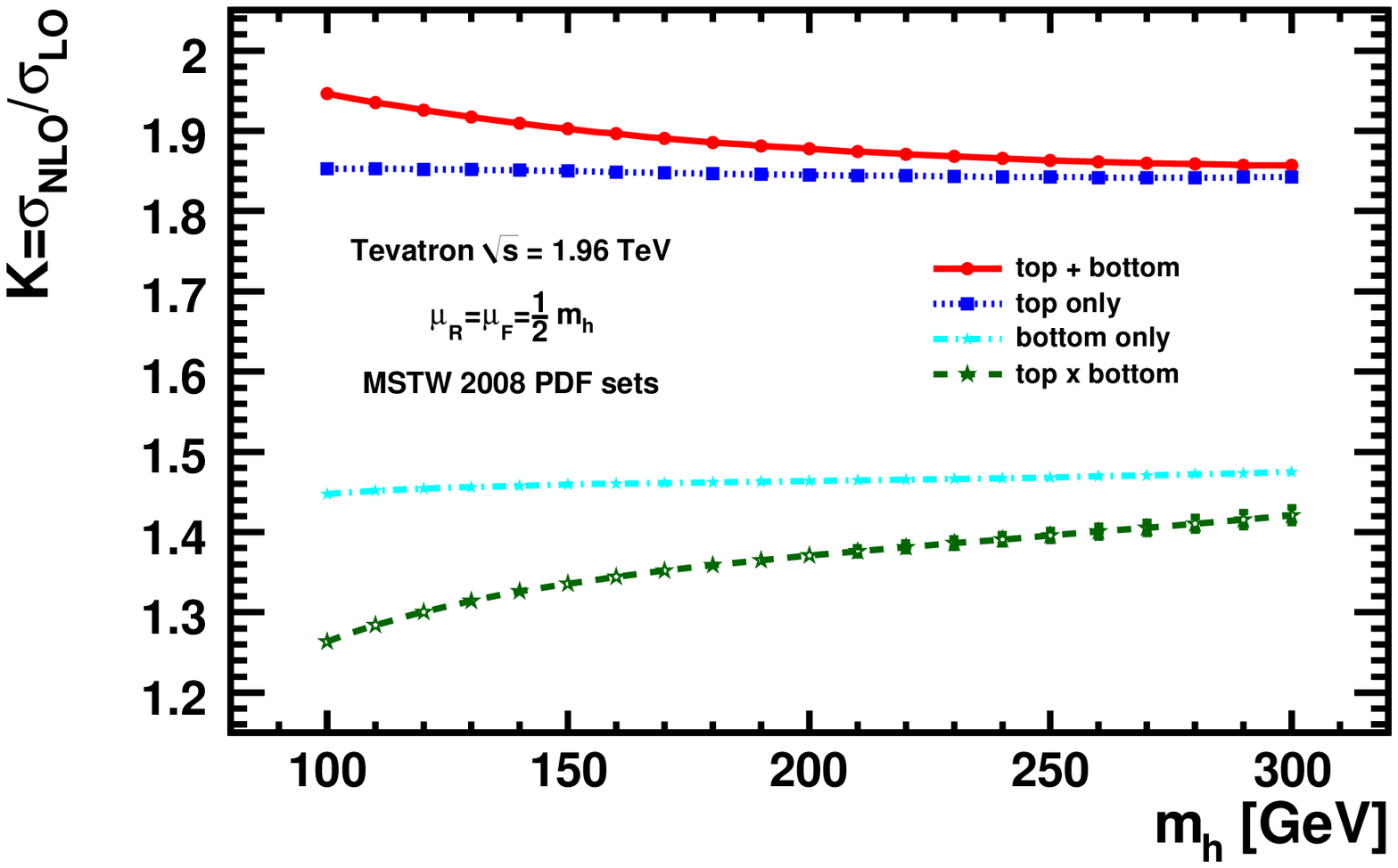}
 \includegraphics[width=0.48\textwidth]{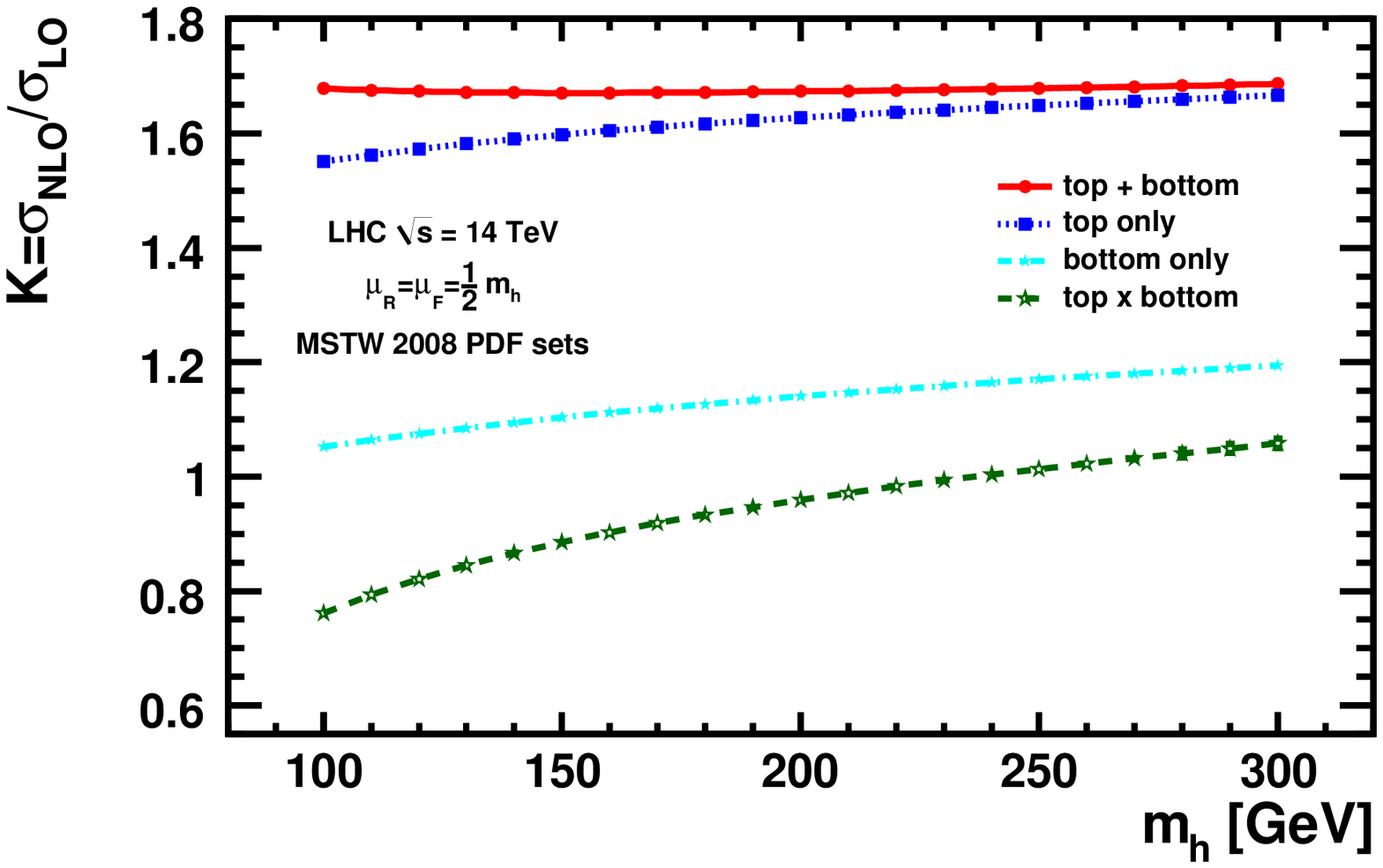}
 \end{center}
  \caption{\label{fig:Kfactor_components}
    NLO K-factors  for  the  top-only, bottom-only, and top-bottom
    interference  contributions. We restrict to a range of $\mh$ where
    top-bottom interference contributions are still sizable.
  }
\end{figure}  
For the Standard Model where  both top and bottom quark loops contribute to the gluon fusion process, we  
study separately the magnitude  of QCD  corrections  for the top-only, bottom-only, and top-bottom interference 
terms. The corresponding K-factors at NLO are plotted in   Fig.~\ref{fig:Kfactor_components}.
While the top-only contributions  receive a large K-factor,  NLO QCD corrections to the 
top-bottom interference and bottom-only terms  are  milder.  As a consequence, the importance of the 
bottom-quark loops is smaller at  NLO than at  LO.

The small NLO QCD  corrections  to the 
top-bottom interference  contribution, which is also a very small fraction of the  top-only contribution, 
suggests that a  more precise evaluation at  NNLO  is not necessary.  The top-only contribution receives 
however large NLO corrections and it requires an evaluation at  NNLO. As  shown in Fig.~\ref{fig:sigma_lhc_diffs}, this 
contribution can be approximated using  Eq.~\ref{eq:hqet} better than  $2\%$ for a  light Higgs boson, and  better than $10\%$ for  a  Higgs boson 
with a mass above the top-pair threshold. It  appears to us,  that the  combination of the NLO cross-section with full dependence on the top and 
bottom quark masses and the NNLO  correction using the approximation of  Eq.~\ref{eq:hqet}  yields a  very precise  estimate of the gluon fusion 
cross-section, where  differences with an NNLO calculation with exact finite
quark mass effects should be quite small.

The results of this section have been extensively cross-checked with {\tt
  HIGLU} and excellent agreement has been found.

\section{Differential cross-sections at NLO with finite quark masses}
\label{sec:differential}

The search for a Higgs boson at hadron colliders is  complicated due to the large cross-sections  of 
background processes.  Sophisticated experimental analyses are required, where  it is   essential  to 
find optimized selection cuts. In addition,  it is  often necessary to perform a detailed probabilistic comparison 
of measured shapes for kinematic distributions with theoretical predictions for the signal and background processes. 
The role of very accurate Monte-Carlo programs which are fully differential is very important  for  these purposes. 

The fully differential NNLO  Monte-Carlo's, 
{\tt FEHiP}~\cite{Anastasiou:2004xq,fehip,Anastasiou:2007mz}  and {\tt HNNLO}~\cite{Catani:2007vq,hnnlo}, are available for the 
gluon fusion Higgs boson production process. Given the complexity of   NNLO computations, these programs employ the approximation 
of  Eq.~\ref{eq:hqet}.  
In some cases, experimental  cuts lead to a significantly smaller scale  variation than in the total cross-section. 
This enhances  the importance  of other  uncertainties, such as the one  due to unaccounted finite quark mass effects.   
A characteristic example  is  the accepted cross-section for  $pp \to H \to WW \to ll\nu \nu$  where a  jet-veto and other cuts 
reduce  the uncertainty due to scale variations by a factor of about two with respect to the  total 
cross-section~\cite{Anastasiou:2007mz,hnnlo,Anastasiou:2009bt}.

With our exact NLO Monte-Carlo {\tt HPro}, we  can correct the predictions of
{\tt FEHiP} and {\tt HNNLO} 
for finite quark mass effects  through NLO. 
{\tt HPro} can be used to compute fully differential cross-sections and distributions  at NLO for
the  Higgs boson and  the final state particles in the two photon and four lepton decays.
{\tt HPro} is merged with  the {\tt FEHiP} NNLO Monte-Carlo and it will be released in a  
forthcoming publication. In this section,  we  illustrate the shapes of  a few kinematic distributions  
with our new {\tt HPro} NLO  Monte-Carlo and compare them with the corresponding predictions in the 
``$m_{\rm top} = \infty$'' approximation.  

\begin{figure}[th]
  \begin{center}
 \includegraphics[width=0.48\textwidth]{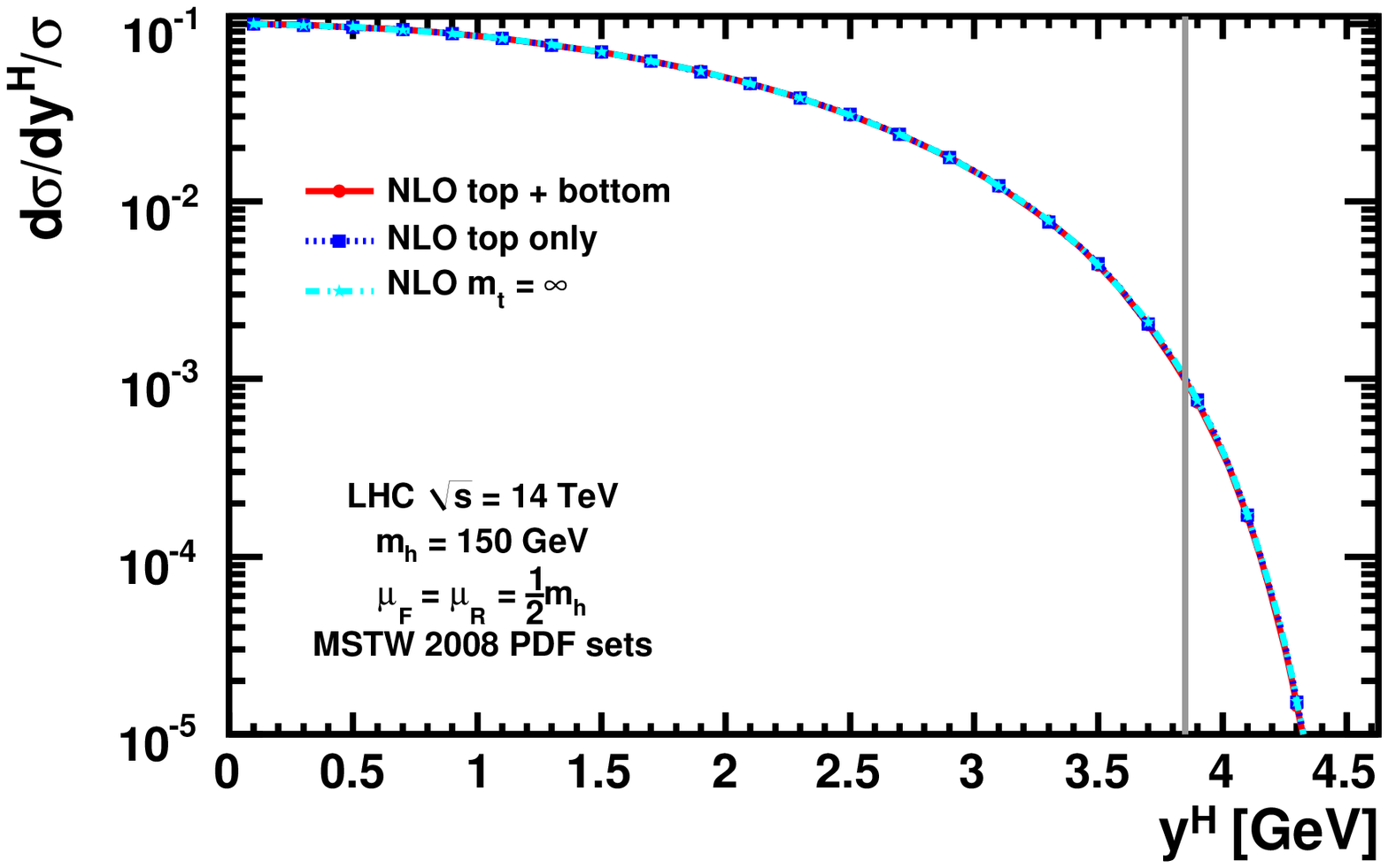}
 \includegraphics[width=0.48\textwidth]{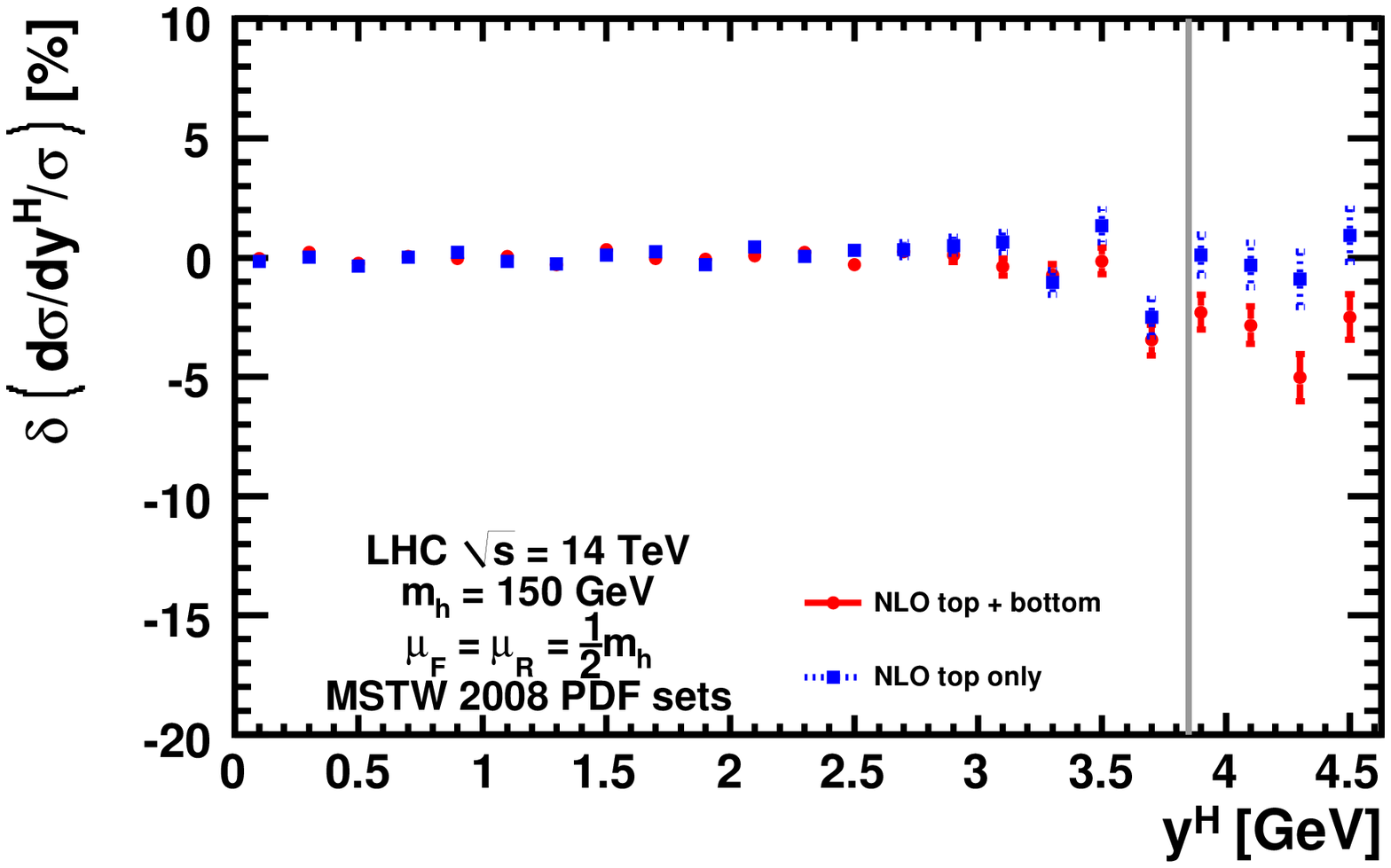} 
 \end{center}
  \caption{\label{fig:rapidityshape_mh150_lhc}
   NLO normalized rapidity distribution at the LHC, $\mh=150\,\GeV$. Finite quark mass effects
   affect the shape of the distributions only in the very high rapidity range,
   where only a tiny fraction of events take place. For $y^{\rm H}$ values
   larger than the one marked by the grey line, less than $10^{-3}$ of the
   total number of events take place.
  }
\end{figure}  
\begin{figure}[th]
  \begin{center}
 \includegraphics[width=0.48\textwidth]{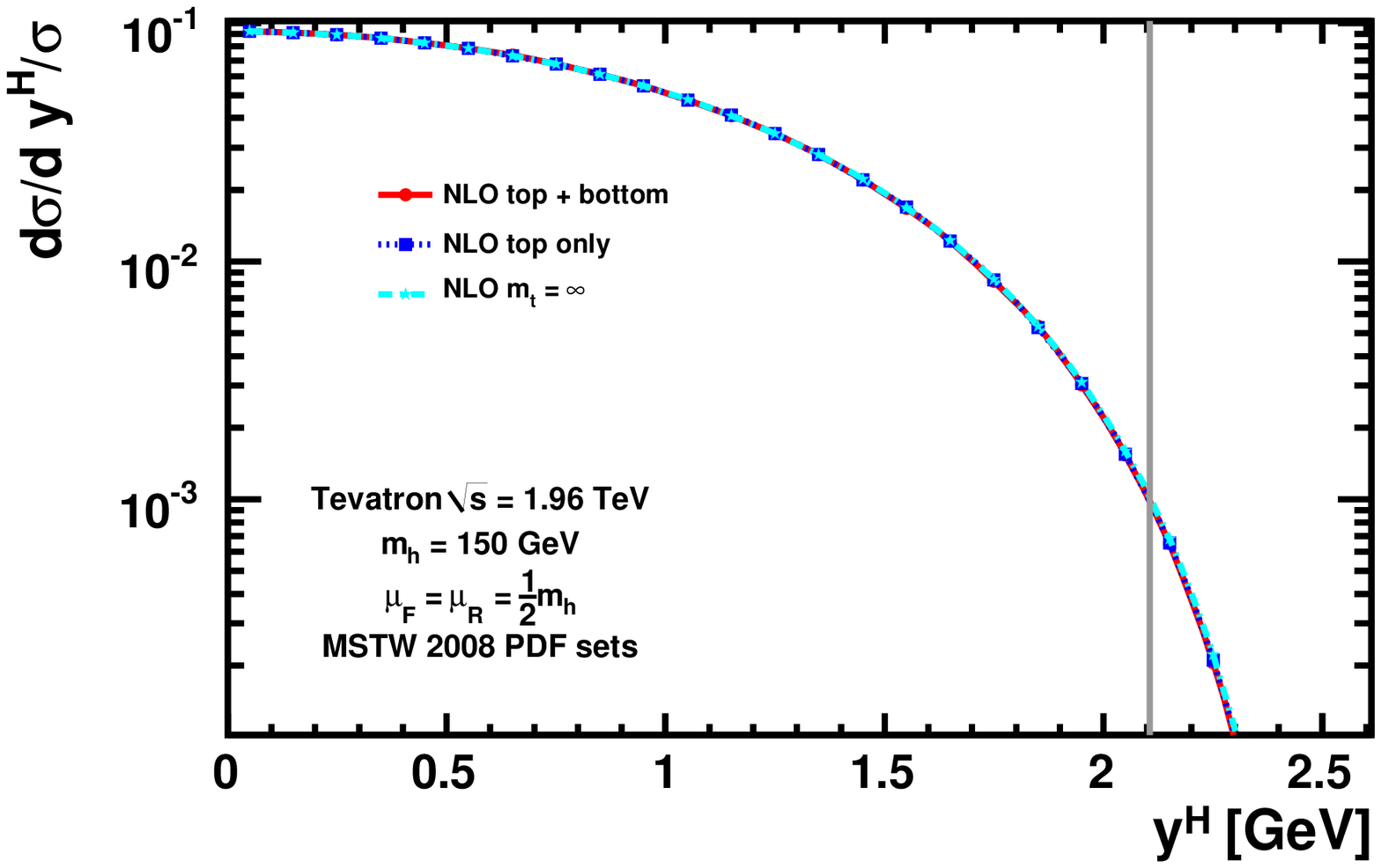} 
 \includegraphics[width=0.48\textwidth]{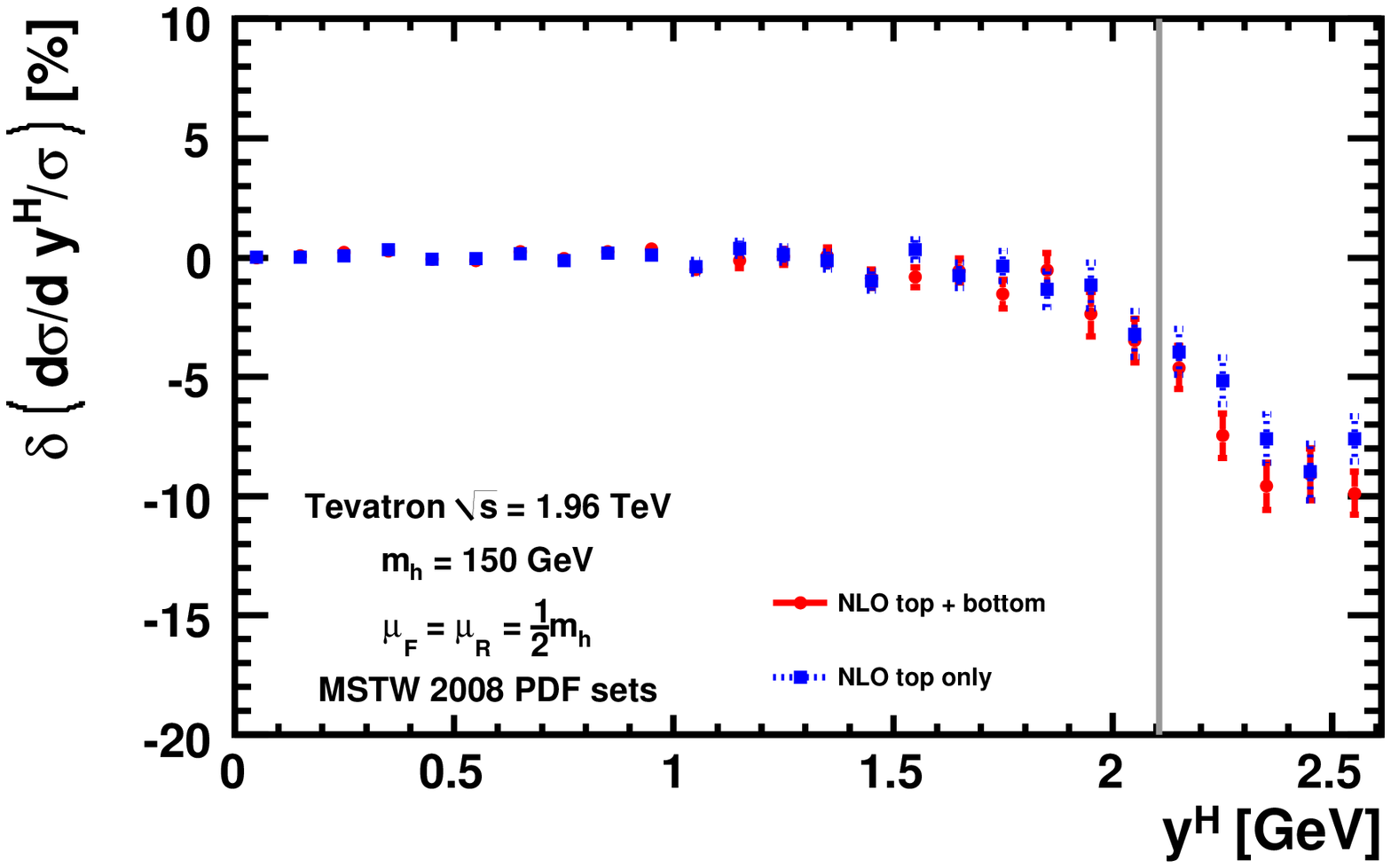} 
 \end{center}
  \caption{\label{fig:rapidityshape_mh150_tev}
   NLO normalized rapidity distribution at the Tevatron, $\mh=150\,\GeV$. Finite quark mass effects
   affect the shape of the distributions only in the very high rapidity range,
   where only a tiny fraction of events take place. For $y^{\rm H}$ values
   larger than the one marked by the grey line, less than $10^{-3}$ of the
   total number of events take place.
  }
\end{figure}  
In Figs.~\ref{fig:rapidityshape_mh150_lhc} and~\ref{fig:rapidityshape_mh150_tev}, we
study the normalized NLO rapidity distribution of  a Higgs boson with mass
$\mh =  150 \, \GeV$ at the
LHC and the Tevatron, respectively. Except for very large rapidities, where almost no events occur, the
distribution is not affected by mass effects. Even in that range the
deviations from the ``$m_{\rm top} = \infty$'' limit are less than $5\%$ at the
LHC and less than $10\%$ at the Tevatron. We conclude that small-$x$ effects are
moderate and without phenomenological consequences in
the case of the rapidity distribution at NLO shedding light on an open question raised in \cite{Marzani:2008az,Marzani:2008ih}.  

Another important differential distribution is the transverse momentum of the Higgs  boson. 
Finite quark mass effects for the $\pt$ distribution have also been studied in earlier 
publications~\cite{Langenegger:2006wu,Baur:1989cm,Ellis:1987xu}, and recently both electroweak and 
finite quark mass corrections were computed and combined~\cite{Keung:2009bs}. 
\begin{figure}[th]
  \begin{center}
 \includegraphics[width=0.48\textwidth]{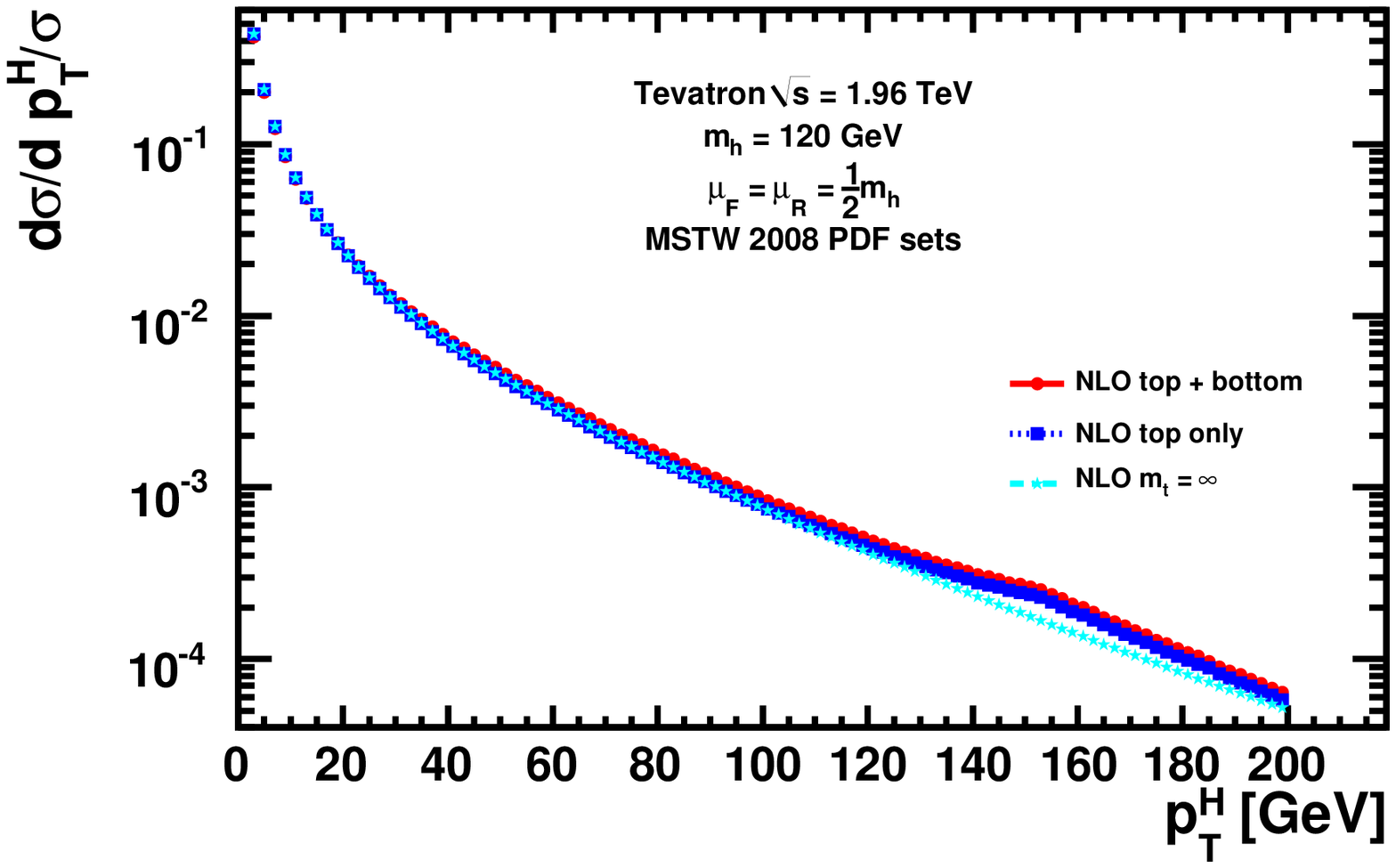}
 \includegraphics[width=0.48\textwidth]{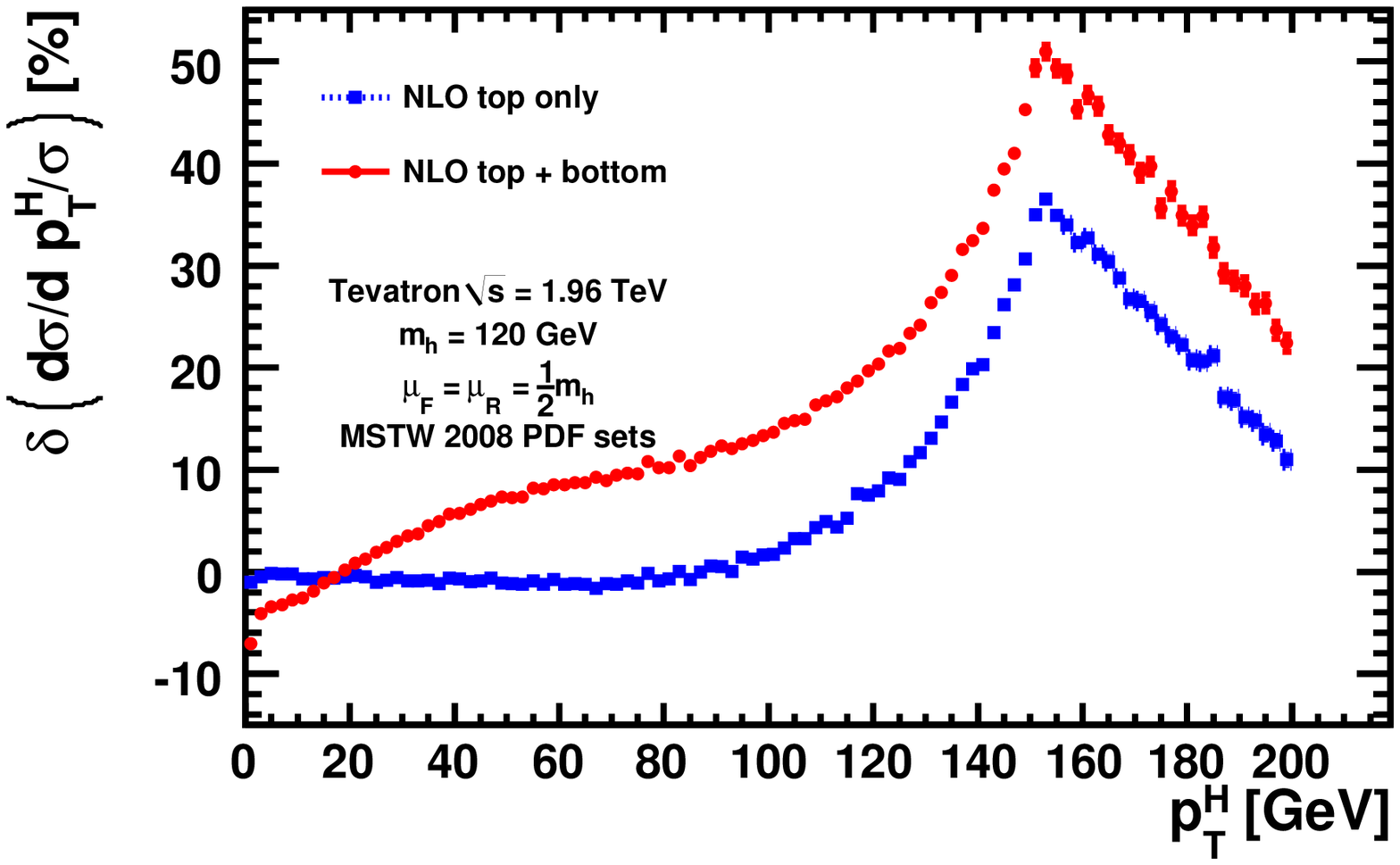}
 \end{center}
  \caption{\label{fig:pt_tev_120}
   Normalized $\pt$ distribution at  Tevatron, $\mh =  120 \, \GeV$. Massive
   corrections are important for large $\pt$. However only a very small
   fraction of events exists in this range. 
 }
\end{figure}  
\begin{figure}[th]
  \begin{center}
 \includegraphics[width=0.48\textwidth]{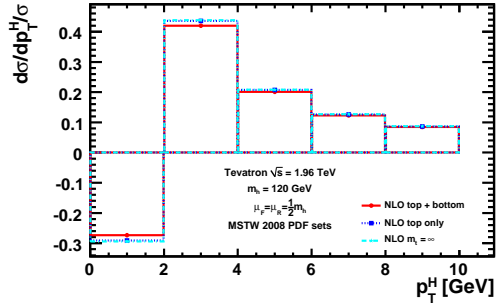}
 \end{center}
  \caption{\label{fig:pt_tev_120_zoom}
   First few bins of $\pt$ distribution at  Tevatron, $\mh =  120 \,
   \GeV$. The values of these bins are unphysical and require resummation.  
  }
\end{figure}  

In Fig.~\ref{fig:pt_tev_120} (left panel) we present the normalized cross-section at the Tevatron in $\pt$ bins  of  $2 \, \GeV$,  for  a Higgs boson with mass 
$\mh = 120 \, \GeV$.  At small values of $\pt$ the bin cross-sections  cannot
be computed accurately in perturbation theory, see Fig.~\ref{fig:pt_tev_120_zoom}, and an all orders  resummation is
required~\cite{Bozzi:2003jy,mcnlo}.  A  meaningful result is  obtained, however, when the bins at low $\pt$ are added  up together.  In  order to study the 
effect of finite quark masses it is more convenient if we demonstrate uncombined low $\pt$  bins. For this purpose, we  present the $\pt$ distribution in the 
approximation of Eq.~\ref{eq:hqet} (cyan), in the ``top-only'' approximation (blue) 
where the  bottom loops are ignored but  the top-loops are evaluated  exactly, and with the complete 
``top-bottom'' mass dependence (red). We have compared our results with the authors of \cite{Keung:2009bs}
and found full agreement within numerical errors.
In the right panel of Fig.~\ref{fig:pt_tev_120} we show the percent deviations of the complete result (red) 
and the ``top-only''  approximation for the normalized $\pt$ distribution from the  approximation of Eq.~\ref{eq:hqet}.  At small $\pt$, there are very small differences 
due to finite quark-mass effects. We observe some  important shape deviations
due  to the effect of  top and bottom quark loops at intermediate $\pt$.  
As it has already been observed in Ref.~\cite{Keung:2009bs} finite quark
effects are   very large at high $\pt$, where the quark production channel
becomes dominant.  Note that additional electroweak
corrections affect the shape considerably \cite{Keung:2009bs}.   
\begin{figure}[th]
  \begin{center}
 \includegraphics[width=0.48\textwidth]{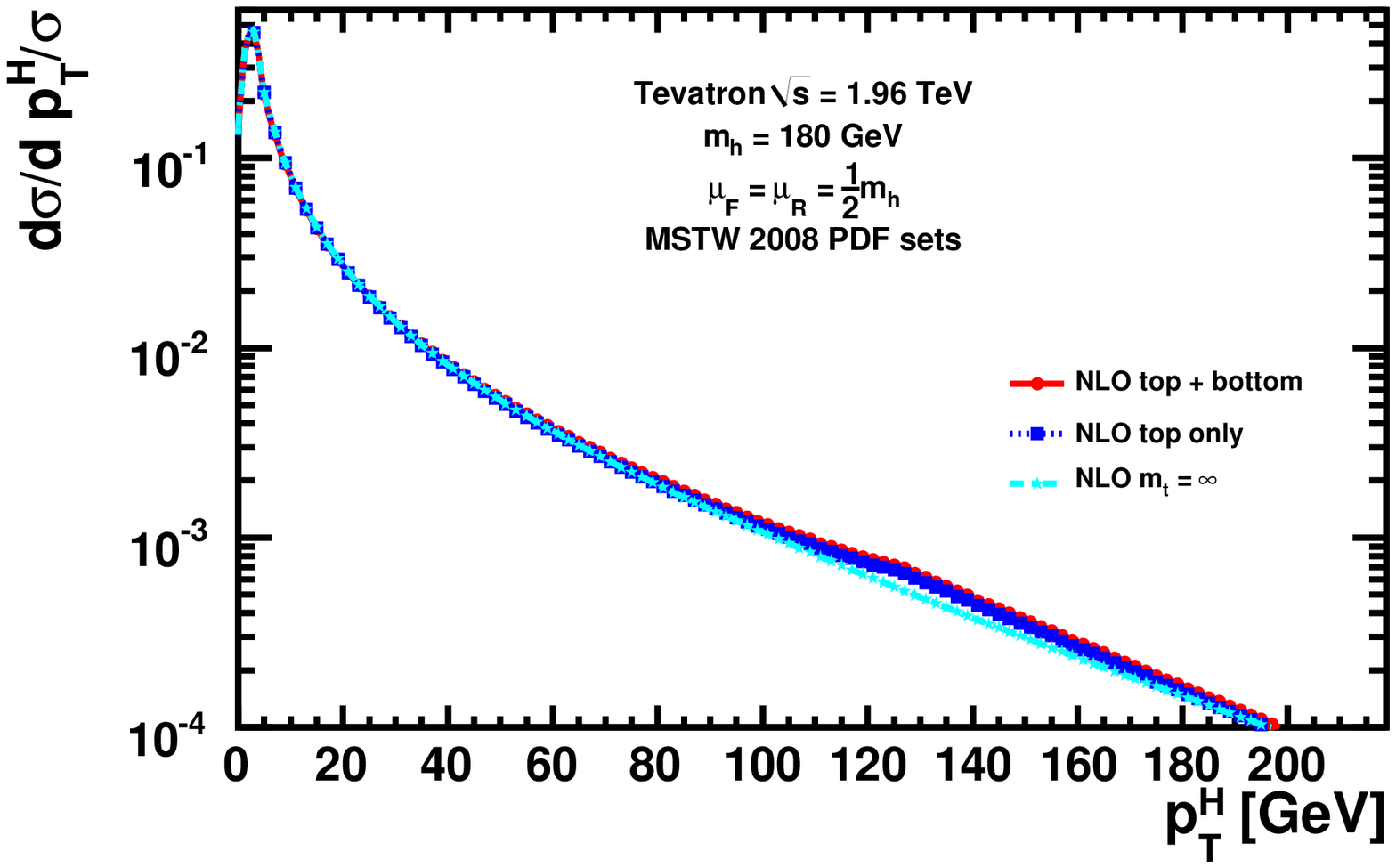}
 \includegraphics[width=0.48\textwidth]{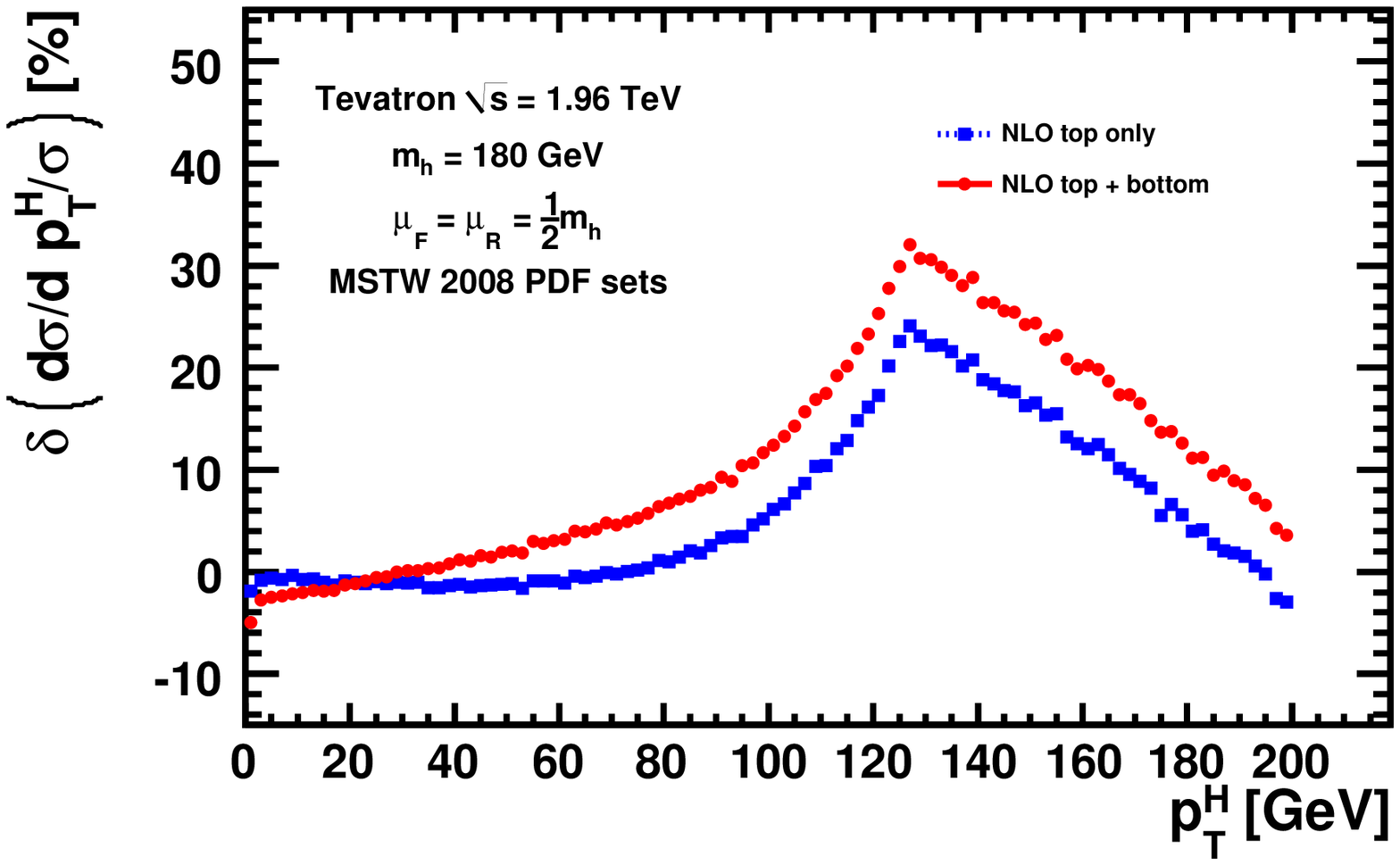}
 \end{center}
  \caption{\label{fig:pt_tev_180}
    Normalized $\pt$ distribution at  Tevatron, $\mh =  180 \, \GeV$. Compared
    to $\mh =  120 \, \GeV$, mass corrections lead to smaller deviations from
    the ``$m_{\rm top}=\infty$'' approximation. The spectrum of the
    bottom-only contribution is much softer (green, left panel). 
 }
\end{figure}  
At  a higher Higgs boson mass value of $\mh = 180 \, \GeV$ (Fig.~\ref{fig:pt_tev_180}) we find an even milder effect at low $\pt$, while  
the magnitude of the deviations at a large $\pt$ is  somewhat reduced but still large. For phenomenological purposes, these  large deviations 
concern a tiny fraction of potential Higgs signal events for both $\mh=120 \,
\GeV$ and $\mh =180 \, \GeV$ mass values. We show in
Fig.~\ref{fig:pt_tev_180} also the ``bottom-only'' contribution and observe
that in this case the $\pt$ spectrum is much softer, as pointed out in
Ref.~\cite{Langenegger:2006wu}. 

\begin{figure}[th]
  \begin{center}
 \includegraphics[width=0.48\textwidth]{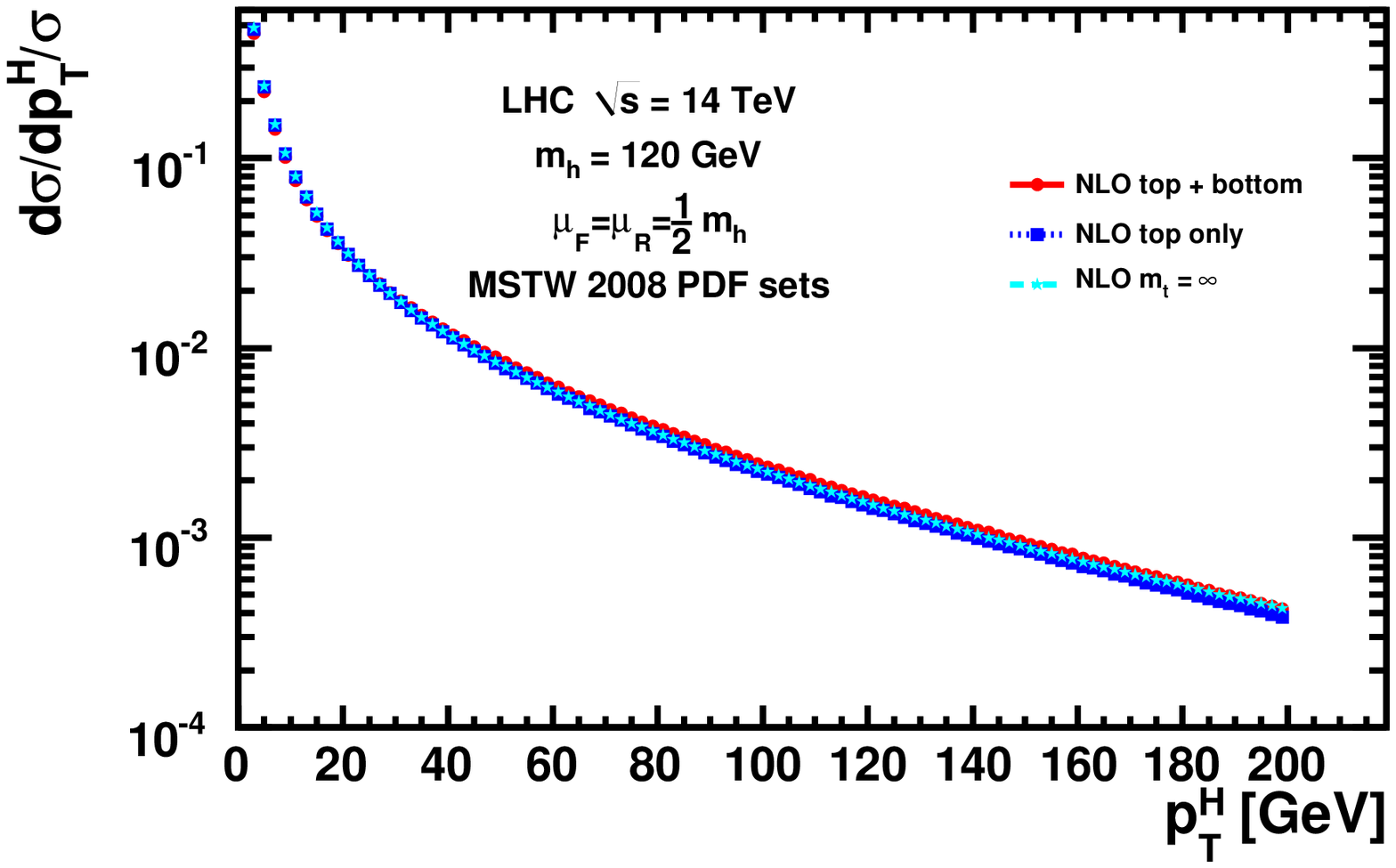}
 \includegraphics[width=0.48\textwidth]{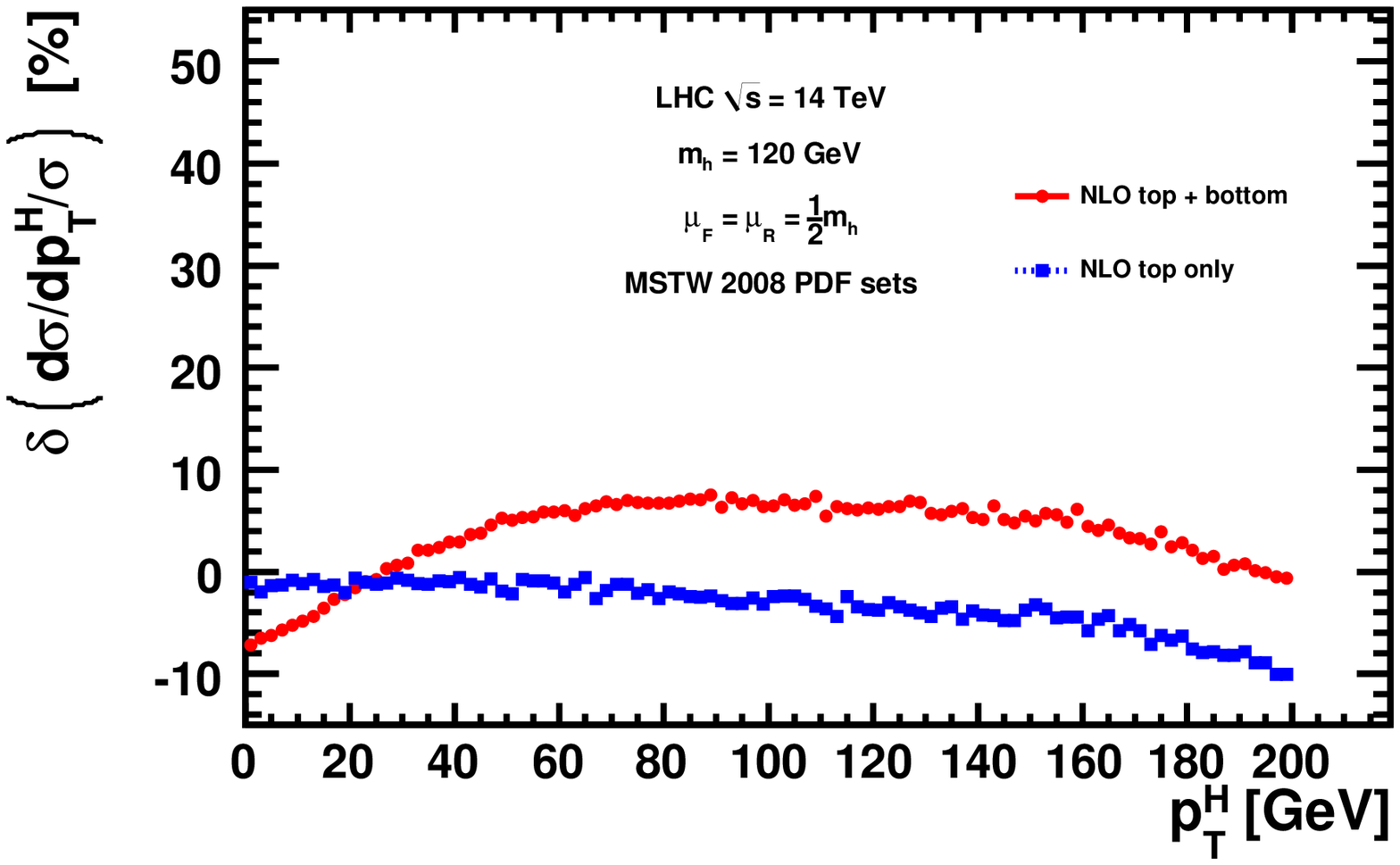}
 \includegraphics[width=0.48\textwidth]{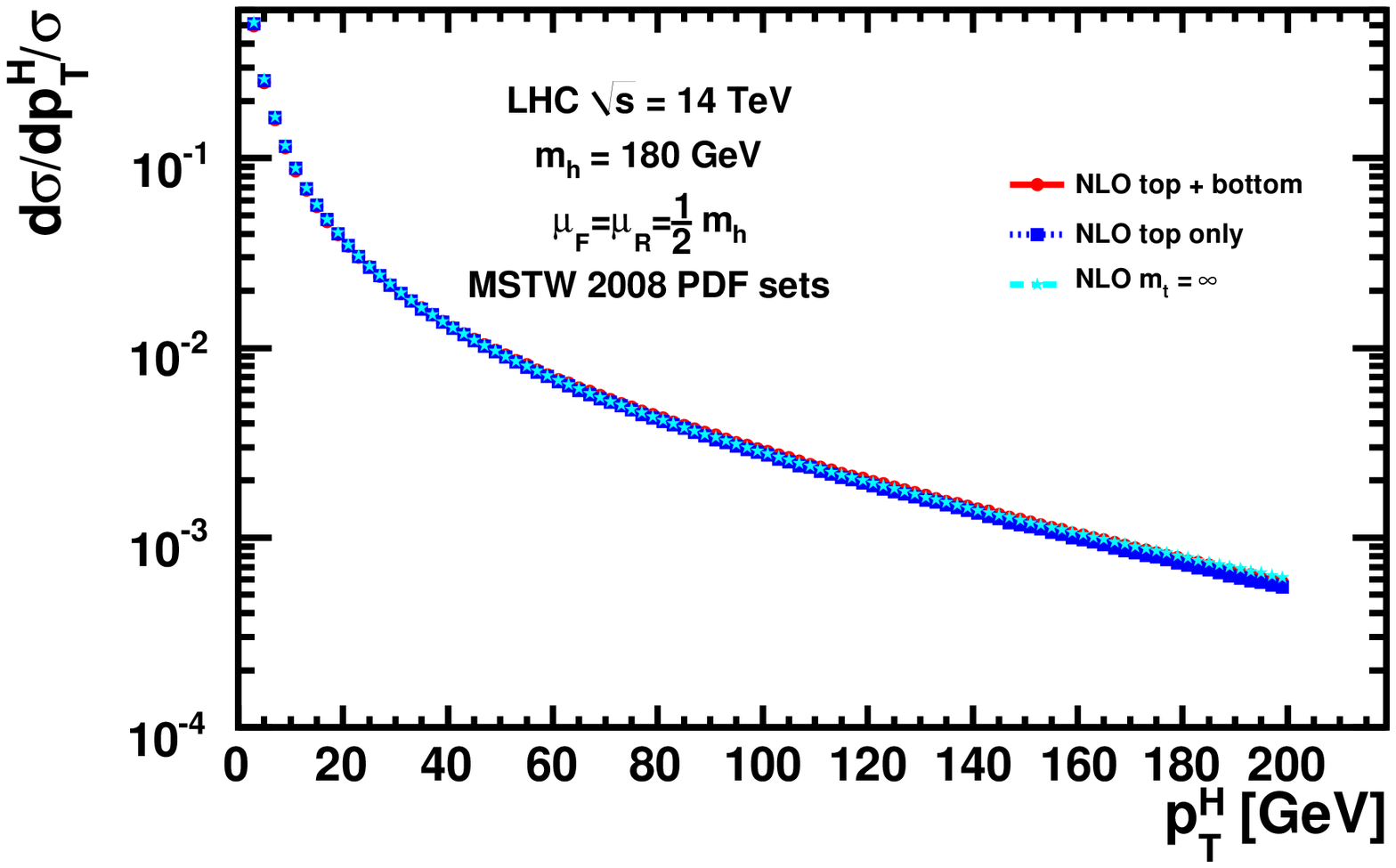}
 \includegraphics[width=0.48\textwidth]{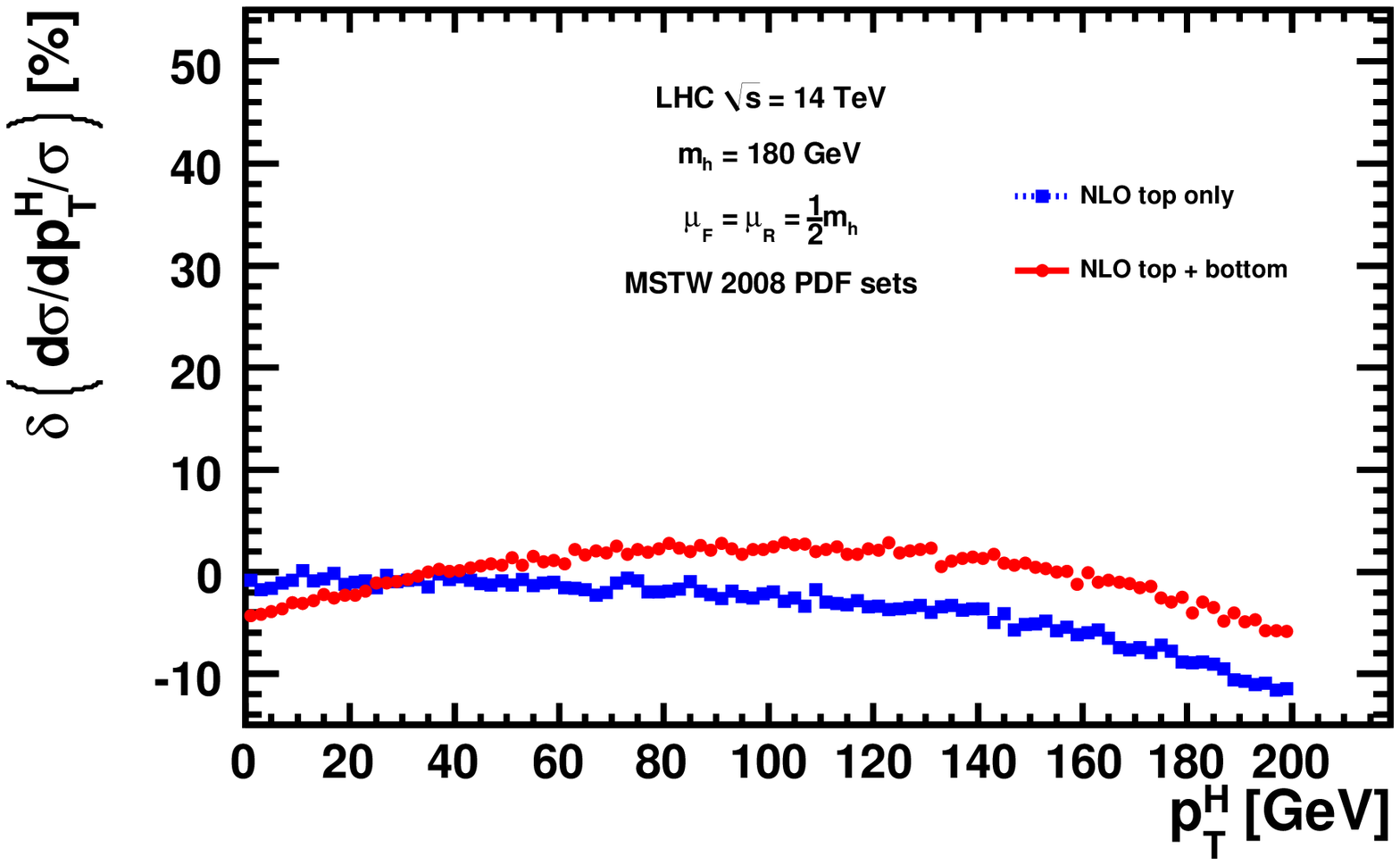}
 \end{center}
  \caption{\label{fig:pt_lhc_180}
   Normalized $\pt$ distribution at LHC for $\mh =  120\, \GeV$ and $\mh= 180
   \, \GeV$. Mass corrections are much more modest than at Tevatron where the
   $q\bar{q}$ channel plays a much bigger r\^ole.
 }
\end{figure}  
At LHC  energy ($14 \,{\rm TeV}$), Fig.~\ref{fig:pt_lhc_180},  we observe significant bottom-loop effects for a light Higgs  boson ($\mh =120 \, \GeV$).  
These are reduced, for a heavier Higgs boson with mass $\mh= 180 \, \GeV$. Shape deviations due to finite quark mass effects  can reach up to 10\% at high $\pt$.  
It is interesting that bottom quark loops for a light Higgs boson change the shape at low $\pt$. As  we  explained, the fixed order $\pt$ spectrum is not 
physical at low $\pt$. However, these  deviations  may also survive after a complete resummation  is performed via the matching procedure.    
\begin{figure}[th]
  \begin{center}
 \includegraphics[width=0.48\textwidth]{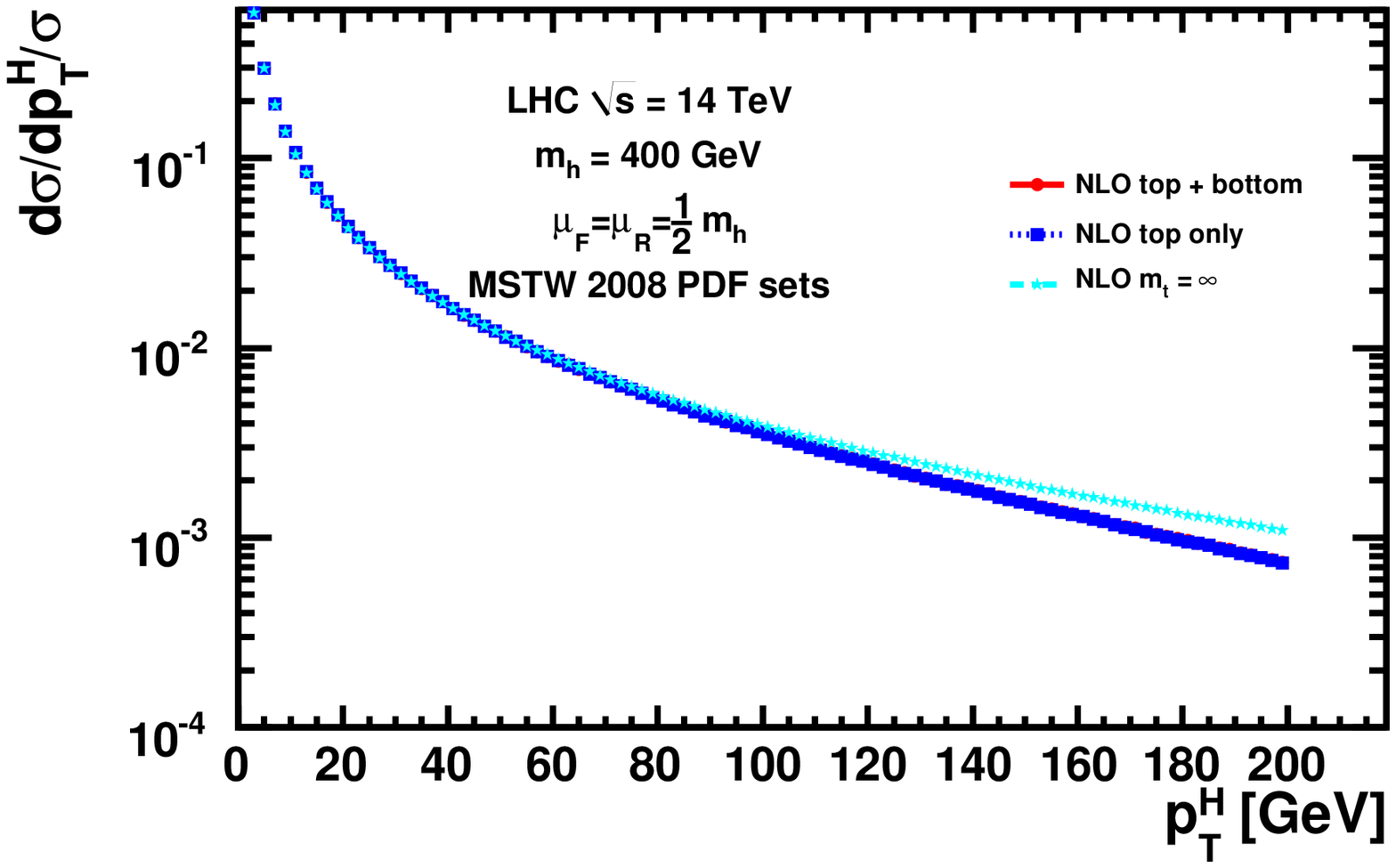}
 \includegraphics[width=0.48\textwidth]{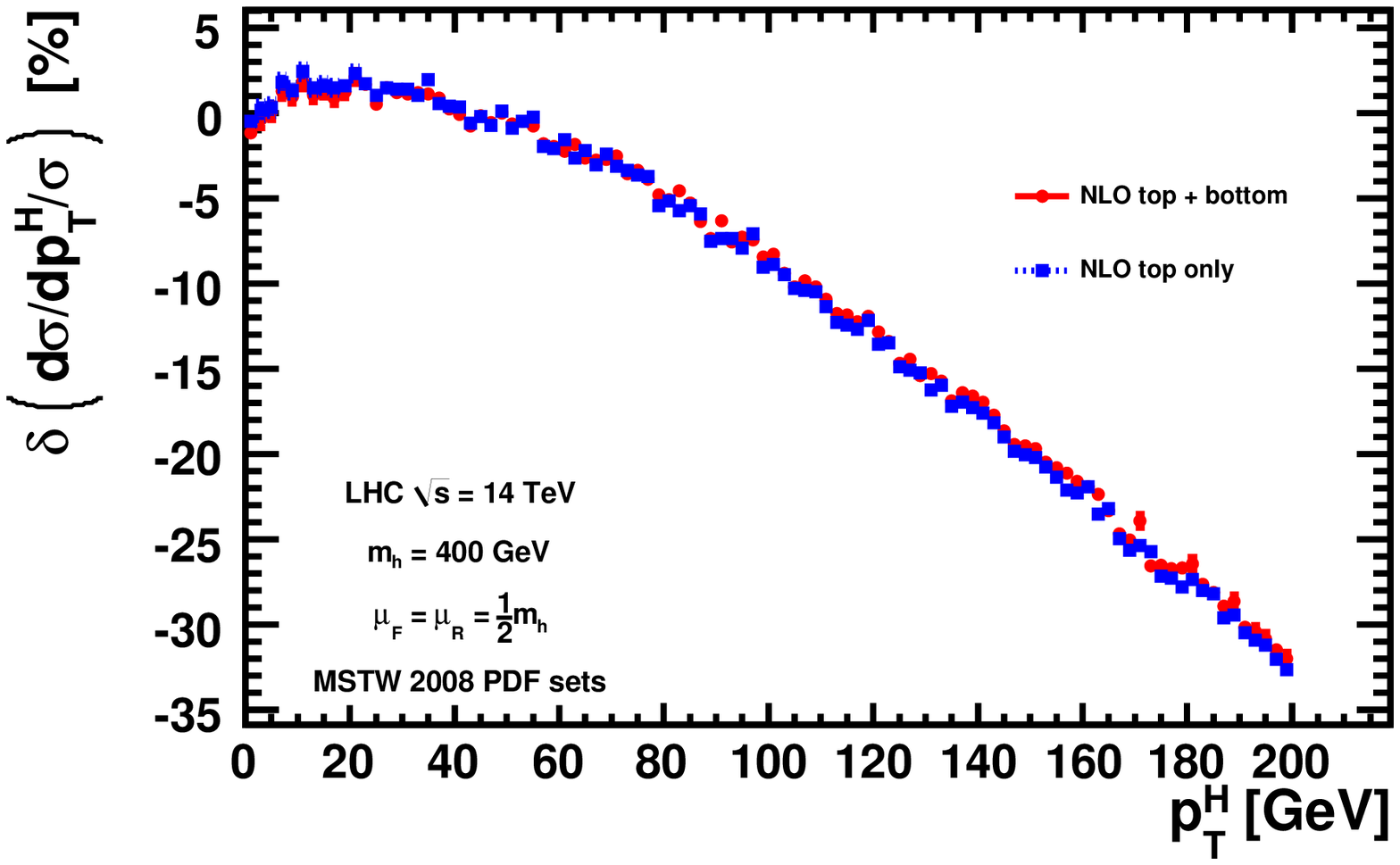}
 \end{center}
  \caption{\label{fig:pt_lhc_400}
   Normalized $\pt$ distribtution at LHC for $\mh =  400 \, \GeV$. Mass
   effects become important as $m_{\rm top}=\infty$ approximation is formally invalid.
 }
\end{figure}  
It is interesting to examine the $\pt$ spectrum for a Higgs mass where the heavy top approximation is formally invalid.  
In Fig.~\ref{fig:pt_lhc_400} we plot the normalized distribution for $\mh =
400 \, \GeV$ at the LHC.  Deviations of the ``top-only'' contributions from
the infinitely heavy top-quark approximation are small for $\pt < 80 \,
\GeV$. At higher $\pt$ the difference increases. 
 
\begin{figure}[th]
  \begin{center}
 \includegraphics[width=0.48\textwidth]{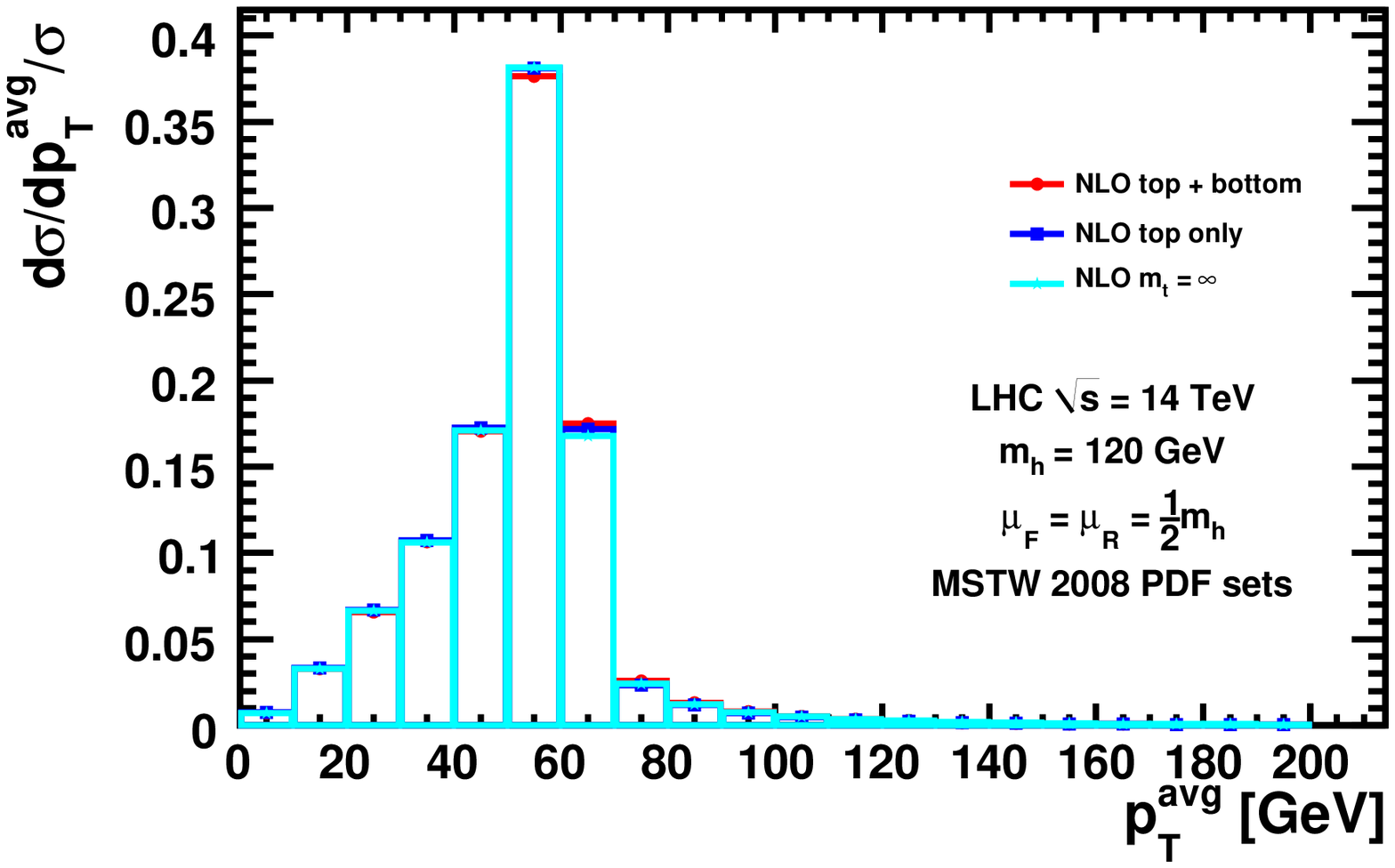}
 \includegraphics[width=0.48\textwidth]{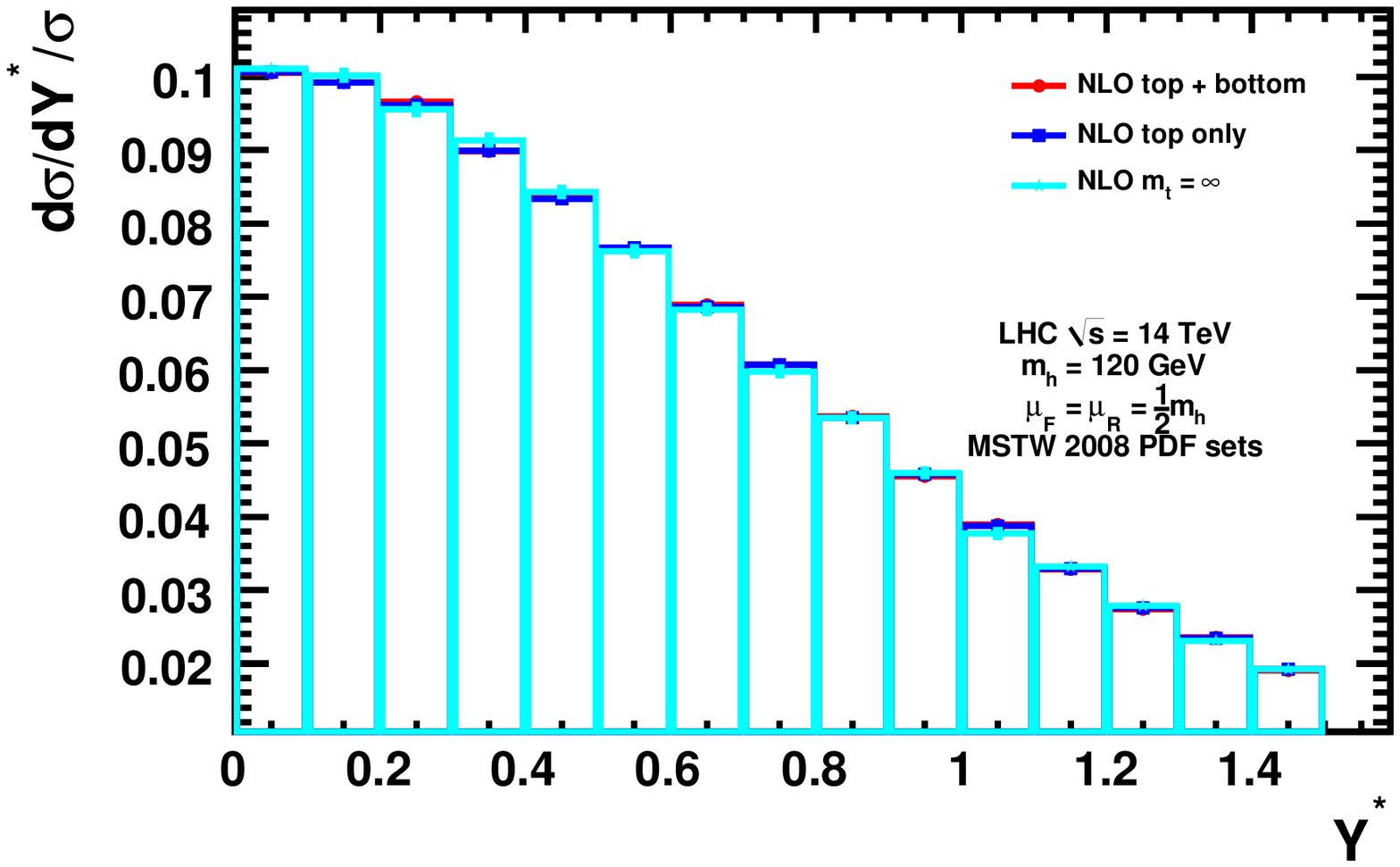}
\end{center}
  \caption{\label{fig:diphoton}
   On the left: Normalized distribution of  the average photon transverse
   momentum, $\pt^{\rm avg}=(\pt^{\gamma_1}+\pt^{\gamma_2})/2$.
   On the right: Normalized distribution of photon pseudorapidity difference, $Y^{*}=|\eta^{\gamma_1}-\eta^{\gamma_2}|/2$.  
   In both plots $\mh =  120 \, \GeV$ and we assume LHC energies.  
 }
\end{figure}  
Finally, we present normalized distributions for Higgs decay final state.  
In Fig.~\ref{fig:diphoton} we present the pseudorapidity difference and average $\pt$ distribution of the two 
photons in the process $pp \to H \to \gamma \gamma$.  Finite quark-mass effects  do not affect these distributions. 
\begin{figure}[th]
  \begin{center}
 \includegraphics[width=0.48\textwidth]{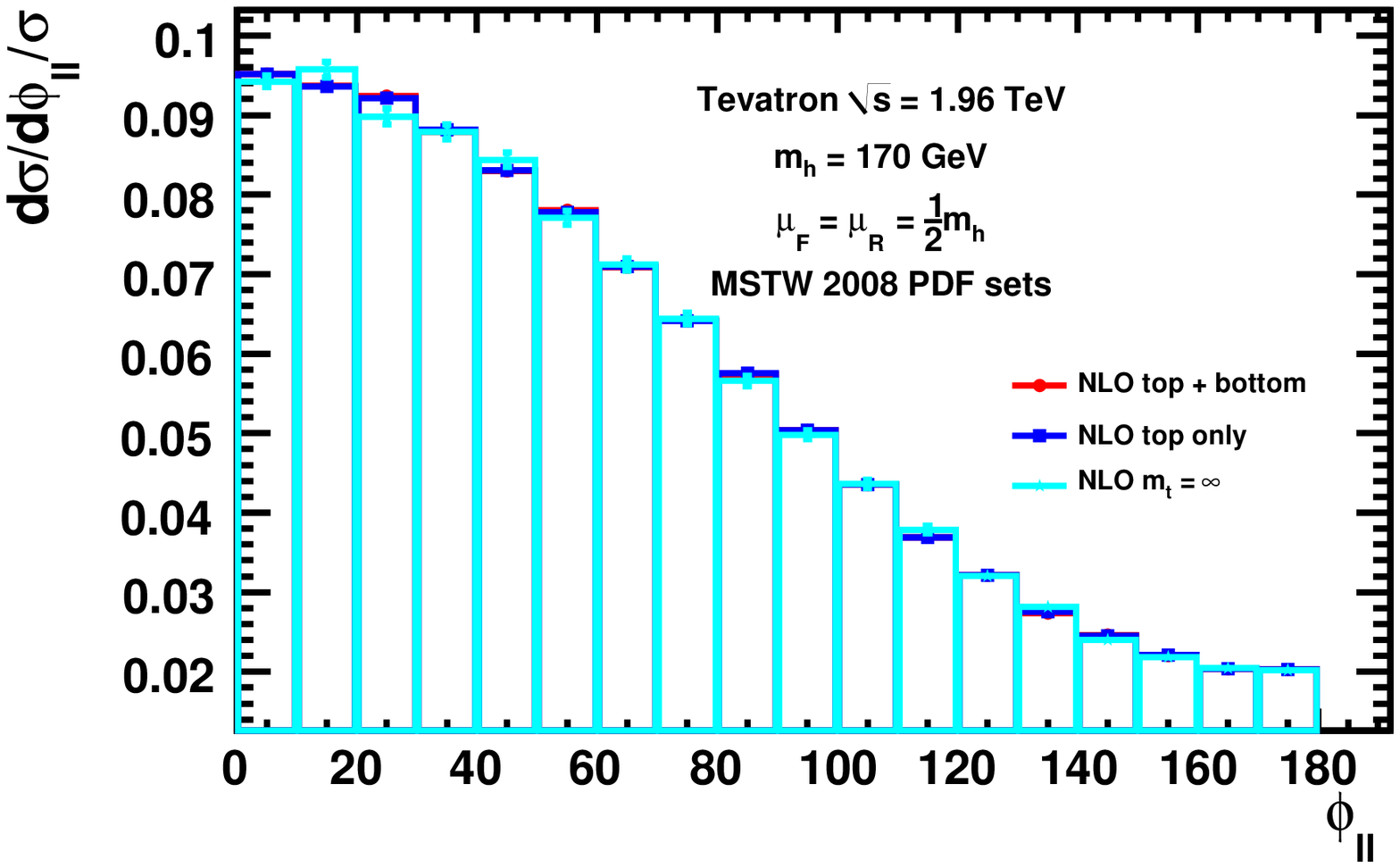}
 \includegraphics[width=0.48\textwidth]{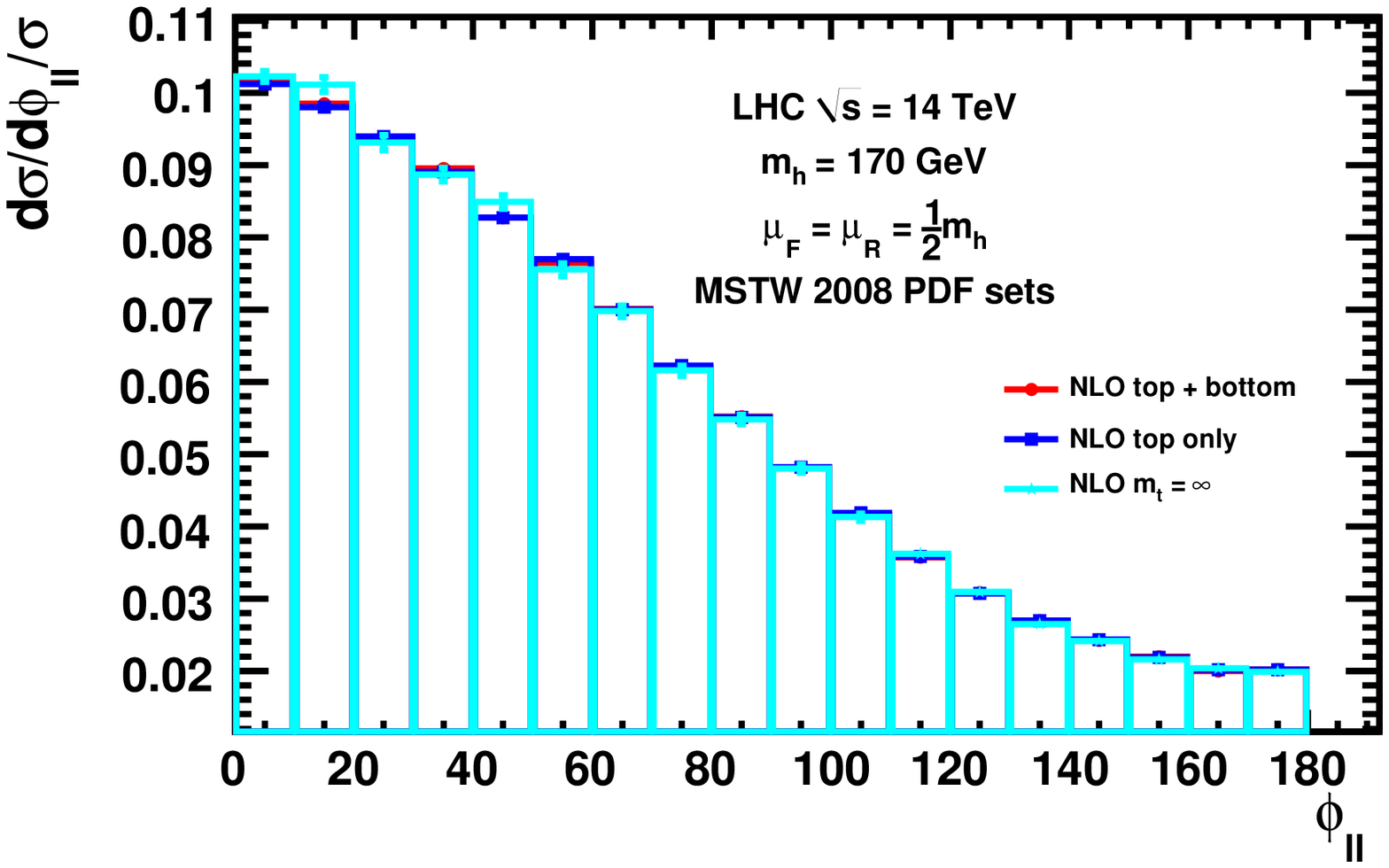}
\end{center}
  \caption{\label{fig:phill}
   Normalized $\phi_{ll}$ distribution for $\mh =  170 \, \GeV$ at
   Tevatron and LHC. $\phi_{ll}$ is the angle in the transverse plane between
   the charged final state leptons, $\phi_{ll}=(p^{l_1}_{\perp}\cdot p^{l_2}_{\perp})/|p^{l_1}_{\perp}||p^{l_2}_{\perp}|$.
 }
\end{figure}  
At higher Higgs boson masses the process $pp \to H \to WW \to ll\nu \nu$ is
dominating and we show  as an example for this decay mode the $\phi_{ll}$
distribution in Fig.~\ref{fig:phill}. $\phi_{ll}$ is the angle in the
transverse plane between the two
charged leptons in the final state and we find again, that the shape is
very well reproduced by the ``$m_{\rm top}=\infty$'' approximation.

In summary, we have found that the shapes of distributions for  leptons and photons from the decay of a Higgs boson are very well approximated 
by Eq.~\ref{eq:hqet}. In addition, accepted cross-sections  after the application of cuts on jets are affected consistently with the expectations  
from the shapes of the Higgs $\pt$ spectra.

\section{Conclusions}
\label{sec:conclusion}

In this paper, we have presented a NLO partonic level  Monte-Carlo program
{\tt HPro}, which computes the 
top and bottom quark mass dependence of differential cross-sections exactly. 
{\tt HPro}  computes accepted cross-sections after selection cuts and kinematic distributions 
for final-state particles in the diphoton and four-lepton decay channels. 

Finite quark mass effects are important and can affect  the precision of NNLO calculations  in the infinite top-quark 
mass approximation. We  can use {\tt HPro} to correct NNLO differential cross-sections  
for finite  quark mass effects. For this purpose, we have interfaced {\tt HPro} with the fully exclusive NNLO Monte-Carlo 
{\tt FEHiP}. The combined program will be  released in a forthcoming publication.  

\section*{Acknowledgements}
We thank Frank Petriello for useful discussions and comparisons. We are 
grateful to Alejandro Daleo for a stimulating collaboration at early stages  
of this work. This work was supported by the Swiss National Science 
Foundation under contract 200021-117873.

\appendix
\section{Real radiation matrix elements}
\label{sec:app_real}
Here we present the results for the matrix elements squared for real radiation processes $gg\to gh$, $qg\to qh$ and
$q\bar{q}\to gh$ in terms of helicity amplitudes. We introduce the following shorthand notation for scalar one-loop integrals, 
\begin{align}
\Bubble(s,m) & =  -\mathrm{i} (4\pi)^2 B_0^{\mathrm{fin}}(s;m) \, , \\
\TrianB(s,m) & = -\mathrm{i} (4\pi)^2 C_0(0,0,s;m,m,m) \, , \\
\TrianA(s,t,m) & = -\mathrm{i} (4\pi)^2 C_0(0,s,t;m,m,m) \, , \\
\Boxx(s,t,m) & = -\mathrm{i} (4\pi)^2 D_0(0,0,0,\mhiggs^2,s,t;m,m,m,m) 
\end{align}
where $B_0$, $C_0$ and $D_0$ are the standard one-loop integrals in the notation of e.g. \cite{Hahn:1998yk}.
$B_0^{\mathrm{fin}}$ is the finite part of the $B_0$ function, i.e. 
\begin{equation}
B_0(s;m) = \frac{\mathrm{i}}{(4\pi)^2} \left(\frac{1}{\eps} + B_0^{\mathrm{fin}}(s;m) \right) \, .
\end{equation}
There exist several publicly available packages for evaluating one-loop integrals \cite{Hahn:1998yk,vanOldenborgh:1990yc,Ellis:2007qk}. However, we used a private implementation and checked against these packages.  

In the following we present the real radiation matrix elements squared,
averaged over spin and color and divided by the flux factor, in terms of
helicity amplitudes, closely adapting the notation in
\cite{Baur:1989cm}. Note however the different convention regarding helicity labels. The simplest process is $q\bar{q}\to gh$ and the result can be expressed by a
single independent helicity amplitude,
\begin{equation}
\Upsilon^{r}_{q\bar{q}}(\s(1,2),\s(1,3),\s(2,3)) = 
 \frac{\as{}^3  (N_c^2-1)}{16 \pi  N_c^2 \s(1,2)^2} \bigg( \Big| \sum_Q M_{Q}^{q\bar{q}} (\s(1,2),\s(1,3),\s(2,3)) \Big|^2 
+  \Big|\sum_Q M_{Q}^{q\bar{q}} (\s(1,2),\s(2,3),\s(1,3)) \Big|^2 \bigg) \, .
\end{equation}
with 
\begin{equation}
\begin{split}
M_{Q}^{q\bar{q}}(\s(1,2),\s(1,3),\s(2,3))  & =  \frac{ \Lambda_Q  m_Q^2 {\s(2,3)}}{{\s(2,3)}+{\s(1,3)}}\bigg[
2 ({\s(1,3)}+{\s(2,3)}-4 m_Q^2)  \TrianA(\mhiggs^2,{\s(1,2)},m_Q) 
\\& \quad {} 
+ \frac{4 {\s(1,2)} }{{\s(2,3)}+{\s(1,3)}} \left( \Bubble({\s(1,2)},m_Q)-  \Bubble(\mhiggs^2,m_Q)\right)-4 \bigg] \, .
\end{split}
\end{equation}
Here, $m_Q\Lambda_Q$ ($\Lambda_Q=1/v$), is the Higgs-quark-quark coupling and we sum over heavy quarks, $Q=t,b$.

Similarly simple is the result for $qg \to q h$ and $gq \to q h$ sub-processes which is obtained by crossing,
\begin{align}
\Upsilon^{r}_{qg}(\s(1,2),\s(1,3),\s(2,3)) & = - \frac{N_c}{N_c^2-1} \Upsilon^{r}_{q\bar{q}}(\s(2,3),\s(1,3),\s(1,2)) \, , \\
\Upsilon^{r}_{gq}(\s(1,2),\s(1,3),\s(2,3)) & = - \frac{N_c}{N_c^2-1} \Upsilon^{r}_{q\bar{q}}(\s(1,3),\s(1,2),\s(2,3)) \, .
\end{align}
More involved is the expression for the $gg\to gh$ process. However, it has a compact representation in terms of only two independent helicity amplitudes, 
\begin{multline} \label{eq:Ups_gghg2}
\Upsilon_{gg}^{r}(\s(1,2),\s(1,3),\s(2,3))   =   \frac{ \as{}^3  N_c  }{ 8 \pi (N_c^2-1) { { \s(1,2)^2} {{\s(1,3)}}
{{\s(2,3)}}}  }
   \bigg( \Big|\sum_Q M_{Q;++-}^{gg} (\s(1,2),\s(1,3),\s(2,3)) \Big|^2 \\
   + \Big|\sum_Q M_{Q;++-}^{gg} (\s(1,3),\s(1,2),\s(2,3)) \Big|^2 
 + \Big|\sum_Q M_{Q;++-}^{gg} (\s(2,3),\s(1,3),\s(1,2)) \Big|^2 
 + \Big|\sum_Q M_{Q;+++}^{gg} (\s(1,2),\s(1,3),\s(2,3)) \Big|^2\bigg)\, .
\end{multline}
The amplitudes appearing in this expression are given by
\begin{equation}
\begin{split} \label{eq:gghgSMppm}
\lefteqn{ M_{Q;++-}^{gg} (\s(1,2),\s(1,3),\s(2,3))  = \Lambda_Q  m_Q^2 \bigg[
  -\frac{4 {\s(1,2)} ({\s(1,2)}^2-{\s(1,3)}
    {\s(2,3)})}{({\s(2,3)}+{\s(1,2)})({\s(1,3)}+{\s(1,2)})} }\\
& \quad {} -\half  \frac{{\s(1,2)} {\s(2,3)}(4 m_Q^2 {\s(1,3)}-{\s(1,2)} {\s(1,3)})}{{\s(1,3)}} \Boxx({\s(1,2)},{\s(2,3)},m_Q) \\
& \quad {} - \half \frac{{\s(1,2)} {\s(1,3)} (4 {\s(2,3)} m_Q^2-{\s(1,2)} {\s(2,3)})}{{\s(2,3)}} \Boxx({\s(1,3)},{\s(1,2)},m_Q)\\
& \quad {} +\half \frac{{\s(2,3)} {\s(1,3)} (-{\s(1,2)}^2+12 
{\s(1,2)} m_Q^2+4 {\s(1,3)} {\s(2,3)})}{{\s(1,2)}} \Boxx({\s(1,3)},{\s(2,3)},m_Q)\\
&\quad {} - \frac{4{\s(1,3)} (2 {\s(1,2)} {\s(2,3)} + {\s(2,3)}^2)}{({\s(2,3)}+{\s(1,2)})^
2} \Bubble({\s(1,3)},m_Q)- \frac{4 {\s(2,3)} (2 {\s(1,2)} {\s(1,3)}+ {\s(1,3)}^2)}{({\s(1,3)}+{\s(1,2)})^2} \Bubble({\s(2,3)},m_Q)\\
&\quad {} -\frac{2 \s(1,3)\s(2,3)}{({\s(1,3)}+{\s(1,2)})^2({\s(2,3)}+{\s(1,2)})^2} \Big(-4 {\s(1,3)}^2 {\s(1,2)} -2 {\s(2,3)} {\s(1,3)}^2-8 {\s(1,3)} {\s(1,2)} {\s(2,3)} \\
&\quad{} -10 {\s(1,3)} {\s(1,2)}^2-2 {\s(2,3)}^2 {\s(1,3)}
-10 {\s(2,3)} {\s(1,2)}^2-4  {\s(1,2)} {\s(2,3)}^2-8 {\s(1,2)}^3 \Big)
\Bubble(\mhiggs^2,m_Q)\\
&\quad{} +\frac{({\s(1,3)} {\s(1,2)} {\s(2,3)}-4 {\s(1,3)} {\s(2,3)} m_Q^2) ({\s(2,3)}+{\s(1,3)})}{{\s(1,3)}{\s(2,3)}} \TrianA(\mhiggs^2,{\s(1,2)}
,m_Q)\\
&\quad{} +\frac{1}{({\s(2,3)}+{\s(1,2)})} 
(4 {\s(1,3)} {\s(2,3)}^2
-{\s(1,2)} {\s(2,3)}^2
+2 \frac{{\s(2,3)}^3 {\s(1,3)}}{\s(1,2)}
+2 {\s(1,3)} {\s(2,3)} {\s(1,2)}
+{\s(1,2)}^3 \\
&\quad {} + 4 m_Q^2  ( {\s(2,3)}^2  + 2 {\s(1,2)} {\s(2,3)} -{\s(1,2)}^2)
)  \TrianA(\mhiggs^2,{\s(1,3)},m_Q)\\
&\quad{} +\frac{1}{({\s(1,3)}+{\s(1,2)})} 
(4 {\s(2,3)} {\s(1,3)}^2
-{\s(1,2)} {\s(1,3)}^2
+2 \frac{{\s(1,3)}^3 {\s(2,3)}}{\s(1,2)}
+2 {\s(2,3)} {\s(1,3)} {\s(1,2)}
+{\s(1,2)}^3 \\
&\quad {} + 4 m_Q^2  ( {\s(1,3)}^2  + 2 {\s(1,2)} {\s(1,3)} -{\s(1,2)}^2)
)  \TrianA(\mhiggs^2,{\s(2,3)},m_Q)\\
&\quad{} -2 \frac{{\s(1,3)}^2 {\s(2,3)}}{{\s(1,2)}} \TrianB({\s(1,3)},m_Q)-2 \frac{{\s(2,3)}^2 {\s(1,3)}}{{\s(1,2)}} \TrianB({\s(2,3)},
m_Q)\bigg] \, 
\end{split}
\end{equation}
and
\begin{equation}
\begin{split}
\lefteqn{ M_{Q;+++}^{gg} (\s(1,2),\s(1,3),\s(2,3))  = \Lambda_Q m_Q^2 \bigg[
 -4 ({\s(1,3)}+ {\s(2,3)}+{\s(1,2)}) }\\
&\quad {}-\half\frac{{\s(2,3)} {\s(1,2)} (4 m_Q^2 {\s(1,3)}-{\s(1,3)} {\s(2,3)}-{\s(1,2)} {\s(1,3)}-{\s(1,3)}^2)}{{\s(1,3)}} 
\Boxx({\s(1,2)},{\s(2,3)},m_Q)  \\
&\quad {} -\half \frac{{\s(1,2)} {\s(1,3)} (4 {\s(2,3)} m_Q^2-{\s(1,2)} {\s(2,3)}-{\s(1,3)} {\s(2,3)}-{\s(2,3)}^2)}{{\s(2,3)}} \Boxx({\s(1,3)},
{\s(1,2)},m_Q)\\
&\quad {} -\half \frac{ {\s(2,3)} {\s(1,3)} (4 {\s(1,2)} m_Q^2-{\s(1,2)} {\s(2,3)}-{\s(1,2)}^2-{\s(1,2)} {\s(1,3)})}{{\s(1,2)}} \Boxx({\s(1,3)},{\s(2,3)},m_Q) \\
&\quad{} + \frac{({\s(1,3)} {\s(1,2)} {\s(2,3)}+{\s(1,3)} {\s(2,3)}^2+{\s(1,3)}^2 {\s(2,3)}-4 {\s(1,3)} {\s(2,3)} m_Q^2) ({\s(2,3)}+{\s(1,3)})}{{\s(2,3)}{\s(1,3)}} \TrianA(\mhiggs^2,{\s(1,2)},m_Q) \\
&\quad {} + \frac{({\s(1,2)} {\s(2,3)}^2+{\s(1,2)}^2 {\s(2,3)}+{\s(1,3)} {\s(1,2)} {\s(2,3)} -4 {\s(1,2)} {\s(2,3)} m_Q^2) ({\s(2,3)}+{\s(1,2)})}{{\s(1,2)}{\s(2,3)}} 
\TrianA(\mhiggs^2,{\s(1,3)},m_Q) \\
& \quad {} +\frac{({\s(1,2)}^2 {\s(1,3)}-4 {\s(1,2)} m_Q^2 {\s(1,3)}+{\s(1,3)} {\s(1,2)} {\s(2,3)}+{\s(1,2)} {\s(1,3)}^2) ({\s(1,3)}+{\s(1,2)})}{{\s(1,2)}{\s(1,3)}} \TrianA(\mhiggs^2,{\s(2,3)},m_Q)\bigg] \, .
\end{split}
\end{equation}


\bibliographystyle{JHEP}

\end{document}